\documentclass[twocolumn]{aastex631}
\usepackage{amsfonts,amsmath,amssymb,mathrsfs}
\usepackage{graphicx}
\usepackage{subfigure}
\usepackage{epstopdf}
\usepackage{float}
\usepackage{color}
\usepackage{cancel}
\usepackage{ulem}
\usepackage{threeparttable}
\usepackage{smartref}
\usepackage{makecell}
\usepackage{multirow}
\usepackage{verbatim}
\usepackage{textcomp}
\usepackage{hyperref}
\usepackage[misc]{ifsym} 
\usepackage{changepage}
\usepackage{tabularx}
\usepackage{booktabs}
\usepackage{CJK}

\begin{document}
\begin{CJK*}{UTF8}{gbsn}
\title{Novel Cosmological Joint Constraints in Multidimensional Observables Space with Redshift-free Inferences}
\author[0000-0001-7906-0919]{Wei Hong}
\altaffiliation{These authors contributed equally to this work.}
\affiliation{Institute for Frontiers in Astronomy and Astrophysics, Beijing Normal University, Beijing 102206, China}
\affiliation{Department of Astronomy, Beijing Normal University, Beijing 100875, China}

\author[0000-0003-0167-9345]{Kang Jiao}
\altaffiliation{These authors contributed equally to this work.}
\affiliation{Institute for Frontiers in Astronomy and Astrophysics, Beijing Normal University, Beijing 102206, China}
\affiliation{Department of Astronomy, Beijing Normal University, Beijing 100875, China}

\author[0000-0002-8429-7088]{Yu-Chen Wang}
\affiliation{Kavli Institute for Astronomy and Astrophysics, Peking University, Beijing 100871, China}
\affiliation{Department of Astronomy, School of Physics, Peking University, Beijing 100871, China}

\author{Tingting Zhang\href{mailto:101101964@sec.edu.cn}{\textrm{\Letter}}}
\affiliation{College of Command and Control Engineering, Army Engineering University, Nanjing 210017, China}

\author[0000-0002-3363-9965]{Tong-Jie Zhang (张同杰)\href{mailto:tjzhang@bnu.edu.cn}{\textrm{\Letter}}}
\affiliation{Institute for Frontiers in Astronomy and Astrophysics, Beijing Normal University, Beijing 102206, China}
\affiliation{Department of Astronomy, Beijing Normal University, Beijing 100875, China}

\correspondingauthor{Tingting Zhang}
\email{101101964@sec.edu.cn}
\correspondingauthor{Tong-Jie Zhang}
\email{tjzhang@bnu.edu.cn}

\received{2023 May 13}\revised{2023 August 30}\accepted{2023 August 31}\published{published date}

\submitjournal{ApJS}

\begin{abstract}
Cosmology constraints serve as a crucial criterion in discriminating cosmological models. The traditional combined method to constrain the cosmological parameters designates the corresponding theoretical value and observational data as functions of redshift, however, sometimes the redshift cannot be measured directly, or the measurement error is large, or the definition of redshift is controversial. In this paper, we propose a novel joint method to constrain parameters that eliminates the redshift $z$ and makes full use of the multiple observables $\left\lbrace \mathcal{F}_{1,\mathrm{obs}},\mathcal{F}_{2,\mathrm{obs}},\cdots,\mathcal{F}_{M,\mathrm{obs}}\right\rbrace$ spanning in $M$-dimensional joint observables space. Considering the generality of the mathematical form of the cosmological models and the guidance from low to high dimensions, we firstly validate our method in a three-dimensional joint observables space spanned by $H(z)$, $f\sigma_{8}(z)$ and $D_{A}(z)$, where the three coordinates can be considered redshift-free measurements of the same celestial body (or shared-redshift data reconstructed model independently). Our results are consistent with the traditional combined method but with lower errors, yielding $H_0=68.7\pm0.1\mathrm{~km} \mathrm{~s}^{-1}\mathrm{~Mpc}^{-1}$, $\Omega_{m0}=0.289\pm0.003$, $\sigma_{8}=0.82\pm0.01$ and showing alleviated parametric degeneracies to some extent. In principle, our joint constraint method allows an extended form keeping the redshift information as an independent coordinate and can also be readily degraded to the form of a traditional combined method to constrain parameters.
\end{abstract}


\section{Introduction}
\label{introduction}
One of the major challenges in cosmology involves attempting to accurately constrain the parameters of cosmological models. The traditional approach to cosmological parameter inference entails using the observational data atlas $(\mathcal{F}_{obs},z)$, which are modeled with Gaussian
distribution and the likelihood $\mathcal{L}(\boldsymbol{\theta})=P\left(\boldsymbol{\mathcal{F}}_{\mathrm{obs}} \mid \boldsymbol{\theta}\right)$. This likelihood function captures both the information of a cosmology model and the way to interpret the model with observational data. By combining this likelihood with a prior distribution, we can obtain a posterior distribution $P\left( \boldsymbol{\theta}\mid\boldsymbol{\mathcal{F}}_{\mathrm{obs}}\right)$ that reflects our updated understanding of the parameters. Whether a single cosmological observable is used to constrain the parameter independently or multiple observables are combined to infer the parameters, the redshift $z$ is always used to label the observational data and also regraded as an independent variable of the model $\mathcal{F}$ in the $\chi^2$ statistic
\begin{equation}
\chi^2=\sum_{j}^{M}\left(\sum_i^{N_j} \frac{\left[\mathcal{F}_{j}\left(z_i ; \boldsymbol{\theta}^{j}\right)-\mathcal{F}_{j,\mathrm{obs}, i}\right]^2}{\sigma_{j,i}^2}\right),\label{eq1}
\end{equation}
where $\boldsymbol{\theta}^{j}=\left(\theta_1^{j}, \cdots,  \theta_D^{j}\right)^{\mathrm{T}}$ is the $D$-dimensional free parameters vector inferred from the $j$-th sub-atlas $(\mathcal{F}_{j,obs},z)$,  $\mathcal{F}_{j}\left(z_i ; \boldsymbol{\theta}^j\right)$ is the theoretical cosmology parameter at $z_i$ given an idiographic set of parameters $\boldsymbol{\theta}^{j}$ and the error $\sigma_{j,i}^2=(\boldsymbol{\sigma}^{j}_{i})^{T}\cdot\boldsymbol{\sigma}^{j}_{i}$ satisfies a Gaussian distribution with a diagonal covariance matrix corresponding to $(\mathcal{F}_{j,obs},z)$, $N_j$ is the number of observational data for $(\mathcal{F}_{j,obs},z)$, and $M$ is the number of observables $\left\lbrace \mathcal{F}_{1,\mathrm{obs}},\mathcal{F}_{2,\mathrm{obs}},\cdots,\mathcal{F}_{M,\mathrm{obs}}\right\rbrace$ and different observables are independent of each other, are equally weighted. Consequently, the likelihood can be expressed as the product of sub-likelihoods without weighting
\begin{equation}
		\begin{aligned}
			\mathcal{\tilde{L}}(\boldsymbol{\theta})&=\prod_j^MP\left(\boldsymbol{\mathcal{F}}_{j,\mathrm{obs}} \mid \boldsymbol{\theta}^j\right)\\&=\prod_j^M\left(\prod_i^{N_j} \frac{1}{\sqrt{2 \pi \sigma_{j,i}^2}}\right) \exp \left(-\frac{\chi^2}{2}\right).	
		\end{aligned}
\end{equation}
It takes a lot of processing power to compute the likelihood directly on a grid in the parameter space, and multidimensional integrations are required to evaluate the marginal distributions of the posterior. In order to estimate $\mathcal{\tilde{L}}(\boldsymbol{\theta})$,  Markov chain Monte Carlo (MCMC) techniques are often used to sample from it. The pioneering articles in cosmological parameter inference by the MCMC method are \citep{Lewis:2002ah} and \citep{2001CQGra..18.2677C}.

The traditional method considers different observables independently in the $\chi^2$ statistic, such as the theoretical value of $H(z)$ corresponds to the observational Hubble data \citep{Ma:2010mr}, and the theoretical value of $f\sigma_{8}$ corresponds to the observational $f\sigma_{8}$ data, even when combining multiple observables. However, this approach may not fully utilize all the available information in the joint data \citep{Linder:2016xer}. Moreover, the joint observables space only employs $(z,H)\oplus(z,f\sigma_{8})$ rather than $(z,H)\oplus(z,f\sigma_{8})\oplus(H,f\sigma_{8})$, where the elements in parentheses can be considered as space coordinate bases. In addition, we frequently designate the corresponding theoretical value and observational data with redshift, even though there are errors associated with redshift measurements, and we typically do not account for the error of redshift when calculating the $\chi^2$, which may lead to a potentially dangerous consequence: due to the error of redshift, the theoretical value of $\mathcal{F}_{j}\left(z_i ; \boldsymbol{\theta}^j\right)$ at redshift $z_i$ is possibly not the actual theoretical value of $\mathcal{F}_{j}\left(z_i ; \boldsymbol{\theta}^j\right)$ corresponding to the observational value at this redshift, which can aggravate the deviation of the results of constraining parameters. In addition, some observations, such as gravitational waves, fast radio bursts, and quasars, cannot be directly observed for redshift or the redshift is ambiguous. 

The theory of gravitational waves (GWs) is a complex and interdisciplinary topic that merges concepts from general relativity, field theory, astrophysics, and cosmology. Since the first GW detection from a binary black hole merger \citep{2016PhRvL.116f1102A}, the science of gravitational waves is developing rapidly. The prerequisite for obtaining redshift by direct observation of GWs is to find the corresponding electromagnetic counterpart. Fortunately, the redshift value of a particular GWs can be obtained indirectly: firstly, the luminosity distance can be obtained by matching the GW waveform. Secondly, the angular diameter distance can be determined by the gravitational lensing effect of GWs \citep{2020MNRAS.492.3359C,2021arXiv211103634T}. Finally, the redshift value can be calculated by combining the two. However, the redshift obtained with this method may be affected by the computational model \citep{Creighton:2011zz,Bailes:2021tot}. 

Fast radio bursts (FRBs), milliseconds-duration radio bursts that originate predominantly from cosmological distances, are detected for the first time in the Small Magellanic Cloud \citep{2007Sci...318..777L}. For the majority of fast radio bursts without reliable redshift measurements, the redshift can be estimated using the measured dispersion measure (DM) value according to the $DM-z$ relation \citep{Macquart:2020lln}. Nevertheless, due to the inhomogeneity of the intergalactic medium (IGM) caused by large-scale structure, the measured DM may be larger or smaller than the theoretical value for a single fast radio burst, and DM measured at the same redshift may also vary significantly for different lines of sight. Furthermore, the DM contributions from the FRB host galaxies and the immediate media around the FRB sources are unknown and difficult to measure because they are degenerate with $DM_{IGM}$ and $DM_{IGM}$, which is largely uncertain. All of these factors contribute to inaccurate redshifts estimated by the $DM-z$ relation \citep{Zhang:2022uzl,Petroff:2021wug,Xiao:2021omr}. 

Quasars are a subclass of active galactic nuclei, which are extremely luminous galactic cores where gas and dust falling into a supermassive black hole emit electromagnetic radiation across the entire electromagnetic spectrum \citep{1999PASP..111..661S,1995PASP..107..803U,2006ApJS..163....1H}. The redshift of quasars has been controversial topic, with three main perspectives regrading their redshift (a) cosmological redshift, (b) non-cosmological redshift, and (c) both \citep{Ellis:2007ds,1989ApJ...347...29S,1987ApJS...63..615N}. The most prominent characteristic of quasars is their large redshift. According to Hubble's Law, this redshift could be attributed to the retrograde motion of extragalactic objects caused by the expansion of the universe. If this is the case, then quasars must be very distant objects, yet they can still be observed thanks to their extreme luminosity. The surprising and perplexing concentration of so much luminosity in such a tiny radiation region is a mystery. In addition, it has also been observed that the galaxy NGC3384 has a redshift $z=0.003$, and that the six quasars found around it have similar redshifts $z\in[1.11,1.28]$ to each other. These quasars are associated with the galaxy NGC3384. However, the redshift of galaxy NGC3384 is significantly different from that of quasars, indicating that the redshift of the quasar is non-cosmological \citep{1979ApJ...229..489A}. 

Given the challenges and uncertainties associated with obtaining redshift information, it is worthwhile to consider a cosmological parameter constraint method without redshift. In this paper, we propose a redshift-free method for constraining the cosmological parameters. This approach extends the traditional combined method while preserving the available redshift information, allowing for the possibility of errors in redshift measurements. The rest of this paper is as follows: in Section. \ref{Mocop}, we propose a new model of combination cosmology parameters that eliminates the redshift $z$. In Section. \ref{Tsard}, we apply it to a set of carefully simulated data to validate the efficacy of our joint constraint method and assess its capacity to constrain parameters of the cosmological model. In Section. \ref{Rad}, we demonstrate the results of constraint using our joint method and make a discussion of our method. Finally, in Section. \ref{conclusion}, we provide a concise summary of this work. In the appendix, we examine the prospective applications of this methodology to several cosmological domains.
\section{Model of combinations of parameters}
\label{Mocop}
\begin{figure}[htbp]
	\centering
	\includegraphics[width=1\linewidth]{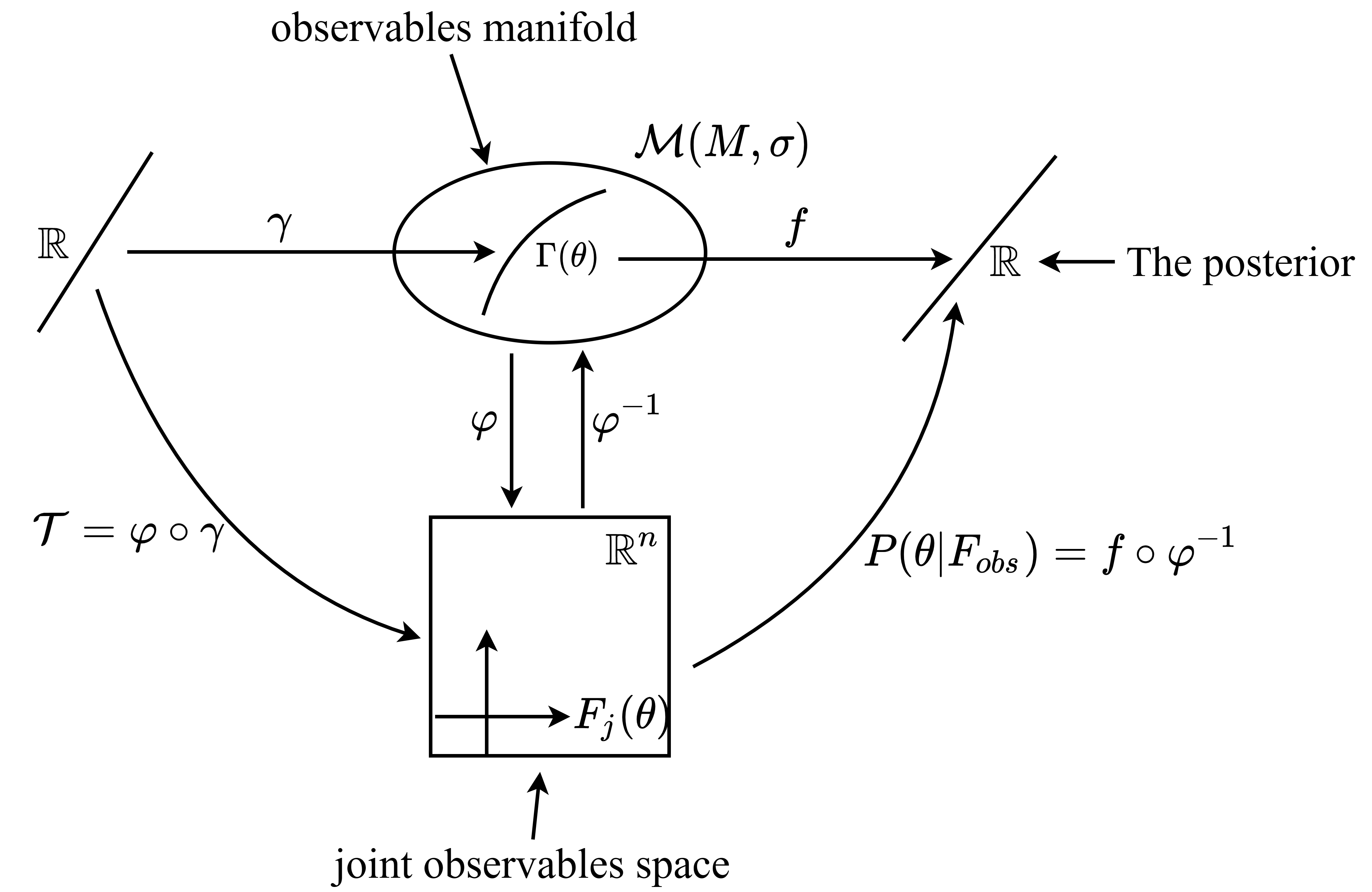}
	\caption{The construction of the observables space and parameter inference.}
	\label{fig:1}
\end{figure}

To break through the aforementioned limitations of existing methods, we propose a novel model for combining cosmological parameters that eliminates the need for redshift $z$ information. This new approach employs the entire joint observables space $\bigoplus_{j,k}^{j\neq k}(\mathcal{F}_{j},\mathcal{F}_{k})$ from the joint theoretical expression $\mathcal{F}_{j}\left(z ; \boldsymbol{\theta}^j\right)$. And we refer to this method as the joint constraint method which is  illustrated in Fig. \ref{fig:1}. All observational data constitute a set $M$ with its topological structure $\sigma=\left\lbrace\varnothing,M\right\rbrace$ becoming a topological manifold $\mathcal{M}$. Firstly, consider an observable curve $\Gamma(\theta)$ on the manifold $\gamma:\mathbb{R}\mapsto \mathcal{M}$ with the parameters $\theta=\left\lbrace H_{0}, \Omega_{m0},\sigma_{8}\dots \right\rbrace$ that can be mapped to an Euclidean space by a composite mapping $\mathcal{T}=\varphi\circ\gamma:\mathbb{R}\mapsto\mathbb{R}$ into the joint observables space. Secondly, with prior knowledge $P(\theta)$, Bayes' theorem yields the posterior as $P\left(\theta \mid H_{\mathrm{obs}}\right)=f\circ\varphi^{-1}$.

\begin{figure}[H]
	\centering
	\includegraphics[width=1\linewidth]{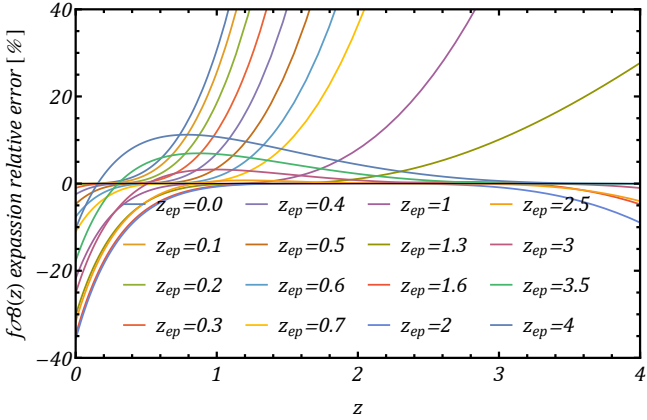}
	\caption{The relative errors in the Taylor expansion of $f\sigma_{8}$ vary with redshift $z$, which are colored corresponding to each expansion point $z_{ep}$ respectively.}
	\label{fig:2}
\end{figure}

For application in practice, we provide the algebraic details for the model presented above. At the beginning, we implicitly determine the redshift by taking the inverse function of the theoretical expression $\mathcal{F}_j\left(z ; \boldsymbol{\theta}^j\right)$. Then, we substitute the inverse function into another theoretical expression $\mathcal{F}_k\left(z ; \boldsymbol{\theta}^k\right)$. For any $\mathcal{F}\left(z ; \boldsymbol{\theta}\right)=f$, we can find the inverse function or redshift of this expression $z=\mathcal{F}^{-1}\left(f ; \boldsymbol{\theta}\right)$, in the range of function segmentation that satisfies bijection, and we can add these segmentations together to find the expressions for the redshift of the entire domain. Since it may be inconvenient to directly obtaining the theoretical expressions of the inverse function, we convert it to a quadratic function by Taylor expansion in a certain range $[z_a,z_b]$ once
\begin{equation}
	\begin{aligned}
\mathcal{F}\left(z ; \boldsymbol{\theta}\right)=&\mathcal{F}\left(z_{ep}; \boldsymbol{\theta}\right)+\frac{\mathcal{F}^{\prime}\left(z_{ep}; \boldsymbol{\theta}\right)}{1 !}\left(z-z_{ep}\right)\\&+\frac{\mathcal{F}''\left(z_{ep}; \boldsymbol{\theta}\right)}{2 !}\left(z-z_{ep}\right)^{2}+O\left(z-z_{ep}\right)^3,
	\end{aligned}
\label{eq3}
\end{equation}
where the function expands at the expanded point $z=z_{ep}$ and $z_{ep} \in [z_a,z_b]$. While the initial and final positions of the solution interval are given, the number and width of the interval can be estimated by using the Lagrangian residue of Taylor expansion as we can set the upper limit of error
\begin{equation}
	P_2(z;z_{ep})=\frac{\mathcal{F}^{(3)}(\xi_{1}; \boldsymbol{\theta}) \cdot\left(z-z_{ep}\right)^{3}}{3!},
\end{equation}
where $\xi_{1}\in\left[\min\left\lbrace z,z_{ep} \right\rbrace,\max\left\lbrace z,z_{ep} \right\rbrace \right]$. Considering that the redshift is to be substituted into another cosmological theoretical expression later, the error of the expansion function should be much smaller than the observation error. Specifically, we put forward a more convenient approach to limiting the upper error limit of expansion $P_{2}(z;z_{ep})\leq\sqrt{\sum_{i}\left( \Delta\theta_{i}/\theta_{i}\right)^2}$ where the $\Delta\theta_{i}$ is the one standard deviation of a cosmological parameter from the Planck catalogue. The Lagrangian residue should be calculated in $\chi^2$ statistic to subtract the error due to higher order expansion terms. Taking $f\sigma_{8}(z)$ as an example, we can calculate the required number of intervals and the width of each interval. The error comparison between the expansion result and the original function is shown in Fig. \ref{fig:2} where $P_{2}(z;z_{ep})\leq\sqrt{\left( \Delta\Omega_{m0}/\Omega_{m0}\right)^2+\left( \Delta\sigma_{8}/\sigma_{8}\right)^2}\approx1.721$ and we set the parameters $\Omega_{m0}=0.315$, $\lambda=1.3$, $\sigma_{8}=0.811$, $\gamma=0.78$, $\beta=1.03$ at different redshift $z$ from 0 to 4.
Once we have the inverse function class $\tilde{z_j}=\mathcal{F}_{j}^{-1}\left(\tilde{f}_j ; \boldsymbol{\theta}^{j}\right)$ of the redshift in $\mathcal{F}_{j}\left(z ; \boldsymbol{\theta}^{j}\right)$ where $\tilde{f}_j=\bigcup_j\left\lbrace f_j\right\rbrace$ is the codomain of $\mathcal{F}_{j}\left(z ; \boldsymbol{\theta}^{j}\right)$, then we can substitute it into another theoretical expression $\mathcal{F}_{k}\left(z ; \boldsymbol{\theta}^{k}\right)$ to eradicate the redshift $\left(\mathcal{F}_{k}\circ\mathcal{F}_{j}^{-1}\right)\left( \tilde{f}_j\right)=\mathcal{F}_{k}\left(\mathcal{F}_{j}^{-1}\left(\tilde{f}_j ; \boldsymbol{\theta}^{j}\right); \boldsymbol{\theta}^{k}\right)$, where $k$ and $j$ are the labels on the function $\mathcal{F}$ which means that the cosmology parameter $\boldsymbol{\theta}$ can be in some cases the same, such as $H(H_0,\Omega_{m0})$ and $D_{A}(H_0,\Omega_{m0})$, but they belong to distinct cosmology functions.  Considering the completeness of the observables space, it is necessary to invert the redshift $\tilde{z_k}$ in $\mathcal{F}_{k}\left(z ; \boldsymbol{\theta}^{k}\right)$ and substitute it into $\mathcal{F}_{j}\left(z ; \boldsymbol{\theta}^{j}\right)$ so that the errors of all observables can be utilized. In order to express the above inverse solution process simply, we introduce the anticommutation notation
\begin{equation}
	\begin{aligned}
			\left\lbrace \mathcal{F}_{j}\left(z ; \boldsymbol{\theta}^{j}\right),\mathcal{F}_{k}\left(z ; \boldsymbol{\theta}^{k}\right)\right\rbrace_{\tilde{f}}&=\mathcal{F}_{j}\left(\mathcal{F}_{k}^{-1}\left(\tilde{f}_k ; \boldsymbol{\theta}^{k}\right) ; \boldsymbol{\theta}^{j}\right)\\&+\mathcal{F}_{k}\left(\mathcal{F}_{j}^{-1}\left(\tilde{f}_j ; \boldsymbol{\theta}^{j}\right) ; \boldsymbol{\theta}^{k}\right).
	\end{aligned}
\label{eq5}
\end{equation}
Similarly, we have the commutation notation
\begin{equation}
	\begin{aligned}
\left[  \mathcal{F}_{j}\left(z ; \boldsymbol{\theta}^{j}\right),\mathcal{F}_{k}\left(z ; \boldsymbol{\theta}^{k}\right)\right] _{\tilde{f}}&=\mathcal{F}_{j}\left(\mathcal{F}_{k}^{-1}\left(\tilde{f}_k ; \boldsymbol{\theta}^{k}\right) ; \boldsymbol{\theta}^{j}\right)\\&-\mathcal{F}_{k}\left(\mathcal{F}_{j}^{-1}\left(\tilde{f}_j ; \boldsymbol{\theta}^{j}\right) ; \boldsymbol{\theta}^{k}\right).
	\end{aligned}
\end{equation}

Therefore, the totally joint observables space can be expressed as
\begin{equation}
	V_{\mathrm{obs}}^{\mathrm{joint}}=\bigoplus_{j,k}^{j\neq k}\left(  \mathcal{F}_{j}\left( \boldsymbol{\theta}^{j}\right),\mathcal{F}_{k}\left(\mathcal{F}_{j}^{-1}\left(\tilde{f}_j ; \boldsymbol{\theta}^{j}\right); \boldsymbol{\theta}^{k}\right)\right),
\end{equation}
where, we let the labels $j$ and $k$ start from 1, scilicet $j=1,2,\cdots,M$. Therefore, the totally joint observables space's model, also regarded as the algebra, can be expressed as
\begin{equation}
	\mathcal{T}\left( V_{\mathrm{obs}}^{\mathrm{joint}}\right) =\bigoplus_{j>k\geqslant 1}\left\lbrace \mathcal{F}_{j}\left(z ; \boldsymbol{\theta}^{j}\right),\mathcal{F}_{k}\left(z ; \boldsymbol{\theta}^{k}\right)\right\rbrace_{\tilde{f}},
	\label{eq8}
\end{equation}
which is the generator of the theoretical cosmology parameters in the $\chi^2$ statistic shown in Eq. (\ref{eq1}). Moreover, the number of the anticommutation notation can be calculated from the expression $\Gamma(M+1)/\Gamma(M-1)$ where $\Gamma(M)$ is the Gamma function. As we know, if there is only one observation, it can not be joint, the number is zero. As we mentioned earlier, this model has a ``model error" $\sigma_{\mathrm{mod},j,k}$ at each $f_k$ due to the existence of the expanded higher-order terms
\begin{equation}
	\begin{aligned}
		\sigma_{\mathrm{mod},j,k}^{2}&=\left[\frac{\partial \mathcal{F}_{j}\left(\mathcal{F}_{k}^{-1}\left(f_k ; \boldsymbol{\theta}^{k}\right) ; \boldsymbol{\theta}^{j}\right)}{\partial \mathcal{F}_{k}^{-1}\left(f_k ; \boldsymbol{\theta}^{k}\right)} \right]^2\sigma_{\mathcal{F}_{k}^{-1}\left(f_k ; \boldsymbol{\theta}^{k}\right)} ^2,
	\end{aligned}
\end{equation}
with
\begin{equation}
	\begin{aligned}
		\sigma_{\mathcal{F}_{k}^{-1}\left(f_k ; \boldsymbol{\theta}^{k}\right)}^2&=\left[\frac{\partial \mathcal{F}_{k}^{-1}\left(f_k ; \boldsymbol{\theta}^{k}\right)}{\partial f_k} \right]^2 \frac{\left(P_2(f_k;\tilde{f}_{k,z_{ep}})\right) ^2}{\left[\mathcal{F}^{\prime}_{k}\left(\xi_{2} ; \boldsymbol{\theta}^{k}\right)\right]^{6}},
	\end{aligned}
\end{equation}
where $\tilde{f}_{k,z_{ep}}$ is the element of the codomain $\tilde{f}_{k}$ corresponding to $z_{ep}$ and $\xi_{2}\in\left[\min\left\lbrace z,z_{ep} \right\rbrace,\max\left\lbrace z,z_{ep} \right\rbrace \right]$. Hence, the total error in Eq. (\ref{eq1}) is changed to $\sigma_{j,k}^2=\sigma_{\mathrm{mod},j,k}^{2}+\sigma_{\mathrm{obs},j,k}^{2}$.

Combining multiple independent observables to infer cosmological parameters is a common approach \citep{1998AJ....116.1009R,2020A&A...641A...6P,2003ApJ...598..102K,2012ApJ...746...85S,2020ApJ...903...83Z}, where composite likelihood is involved. Composite likelihoods are derived by evaluating the likelihoods for subsets of the data and subsequently aggregating these likelihoods under the assumption of independence among the subsets \citep{article-Composite-Likelihood,ae96fe4b-cd0b-3a2b-9a88-5bd548187d36,cfad4d52-869f-35f3-8652-8534d93d28ad,7ba3ad00-48e3-3e1a-bb81-112e59ea3834,23c7e5f7-f5a5-3aad-91bb-4641d35779f8}. Parameter estimates are derived by maximizing the composite likelihoods that arise from the data. There exist two primary motives for employing composite likelihood methodologies in general. Initially, it is computational in nature, as it involves the calculation of likelihoods for subsets of data, which is typically more manageable than calculating the full likelihood for the entire dataset. Furthermore, it circumvents the necessity to incorporate higher-order dependencies in the data modeling process, resulting in inferences that rely solely on the modeling of appropriate marginal or low-dimensional aspects of the data. The inference function exhibits the qualities of likelihood from a misspecified model, regardless of the independence of the components, due to their multiplication. An additional rationale for employing composite likelihood is the consideration of robustness, specifically in scenarios where there may be a potential misrepresentation of the higher-order dimensional distributions. For instance, when dealing with dependent binary data, the utilization of pairwise likelihood obviates the need to select a model for the joint probabilities of triples and quadruples. Likewise, composite likelihood, by its very design, remains resilient against alternative possibilities for these joint probabilities, as long as they align with the modeled joint probabilities of pairs. The concept of robustness being discussed here differs from the notion of robust point estimation. Instead, it aligns more closely with the robustness produced through extended estimating equations. Nevertheless, in the case of high-dimensional models, it remains uncertain which higher-order joint densities are truly consistent with the lower-order marginal densities being modeled. Consequently, investigating the topic of robustness in a comprehensive manner becomes challenging.

In this work, we employ a three-dimensional model $(H(z),f\sigma_{8}(z),D_{A}(z))$ comprised of the Hubble expansion rate $H(z)$ describing the expansion history, the growth rate $f\sigma_{8}(z)$ describing the growth history of the universe and the angular diameter distance $D_{A}(z)$ describing the scale of the object in the universe to verify this method. One can make an inspiration from the expansion history, the growth history, and the scale of the object in the universe with the three-dimensional model. Besides, the theoretical formulations of $H(z)$, $f\sigma_{8}(z)$ and $D_{A}(z)$ are enlightening in mathematics. $H(z)$ is monotonic over the entire domain interval, and the inverse of its redshift can be readily solved. $f\sigma_{8}(z)$ is not monotonic over the entire domain interval with non-integer powers of expression, and the inverse of its redshift should be solved by Taylor expansion. $D_{A}(z)$ is also not monotonic over the entire domain interval with an integral term of expression, and the inverse of its redshift should be solved by Taylor expansion considering complex special functions in real number space. These three expressions contain most of the mathematical forms of cosmologically theoretical expressions. 

Our choice of a three-dimensional model is not only motivated by the fundamental importance of the chosen observables but also by the potential for extension to higher dimensions. By demonstrating the effectiveness of the proposed method on a three-dimensional model, we provide a foundation for its application to more complex higher-dimensional models.

\section{The Simulated and Reconstructed Data}
\label{Tsard}
\begin{figure*}[ht!]
	\centering
	\subfigure[]{
		\includegraphics[width=0.3\linewidth]{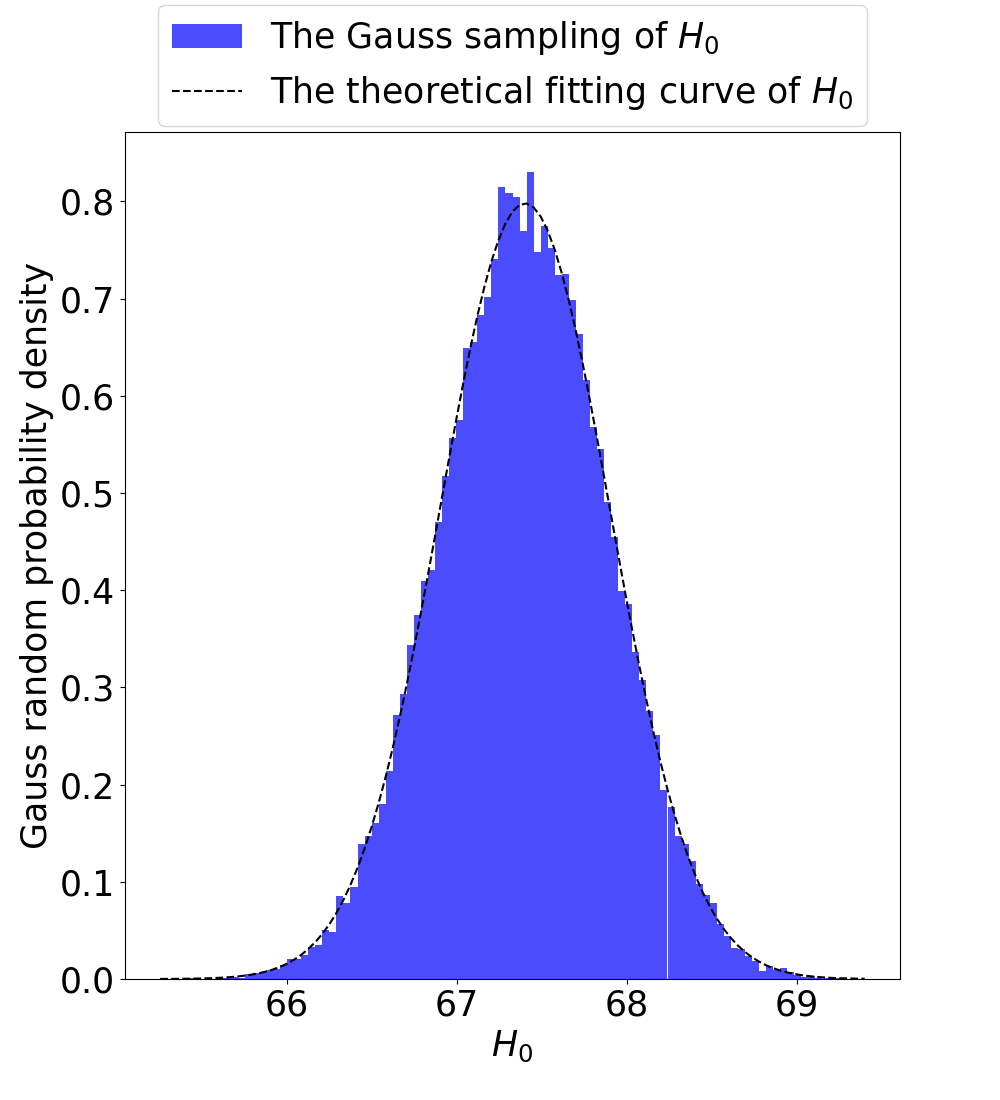}
	}
	\subfigure[]{
		\includegraphics[width=0.3\linewidth]{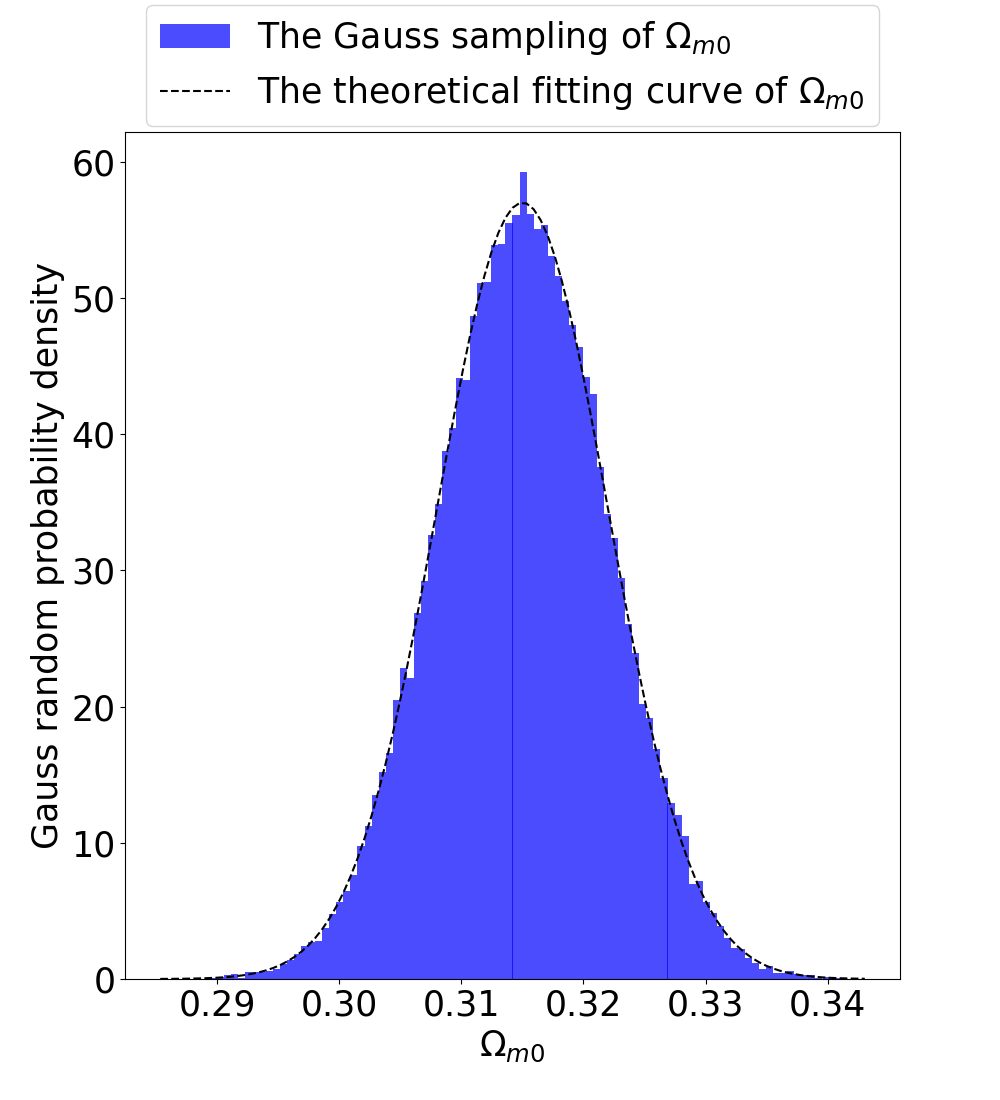}
	}
	\subfigure[]{
		\includegraphics[width=0.3\linewidth]{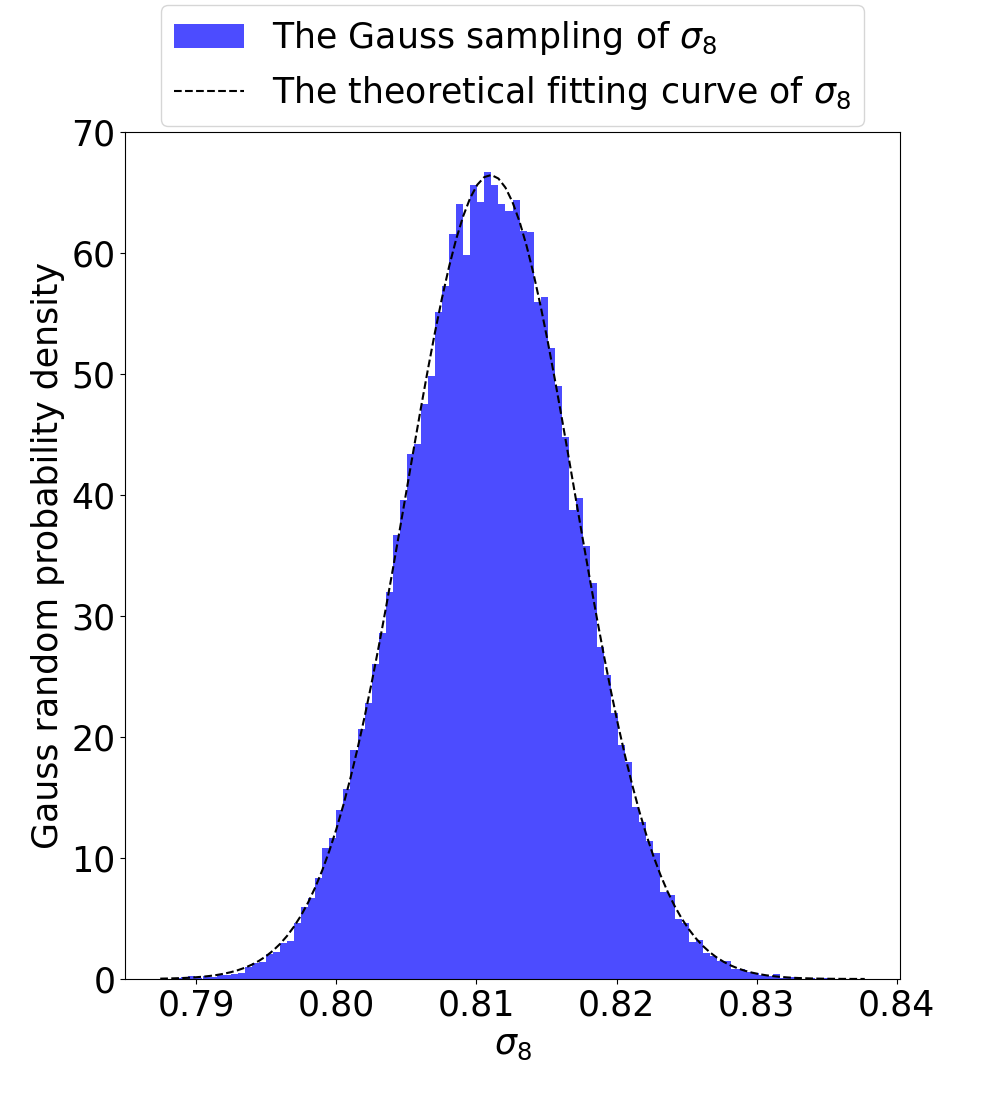}
	}
	\subfigure[]{
		\includegraphics[width=0.3\linewidth]{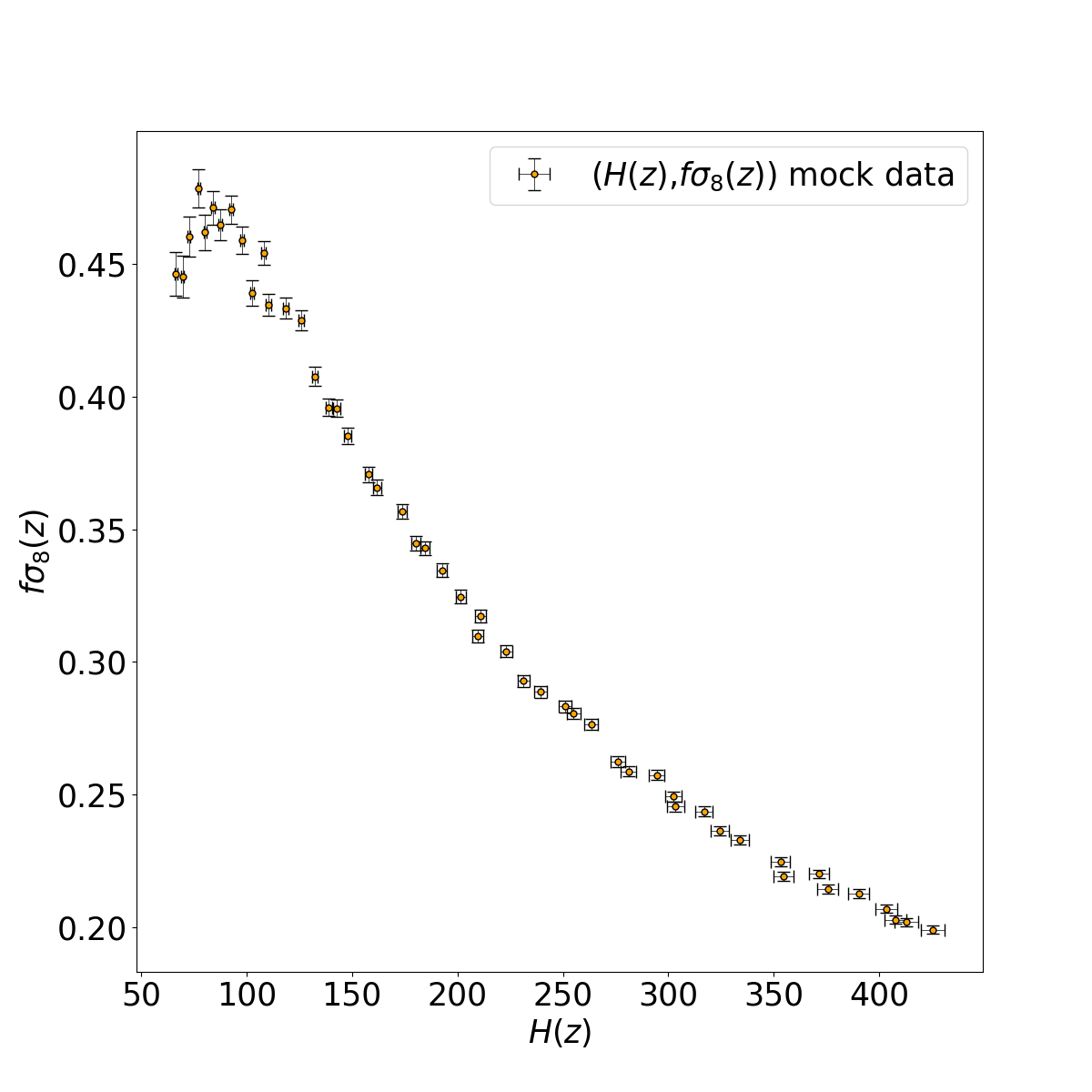}
	}
	\subfigure[]{
		\includegraphics[width=0.3\linewidth]{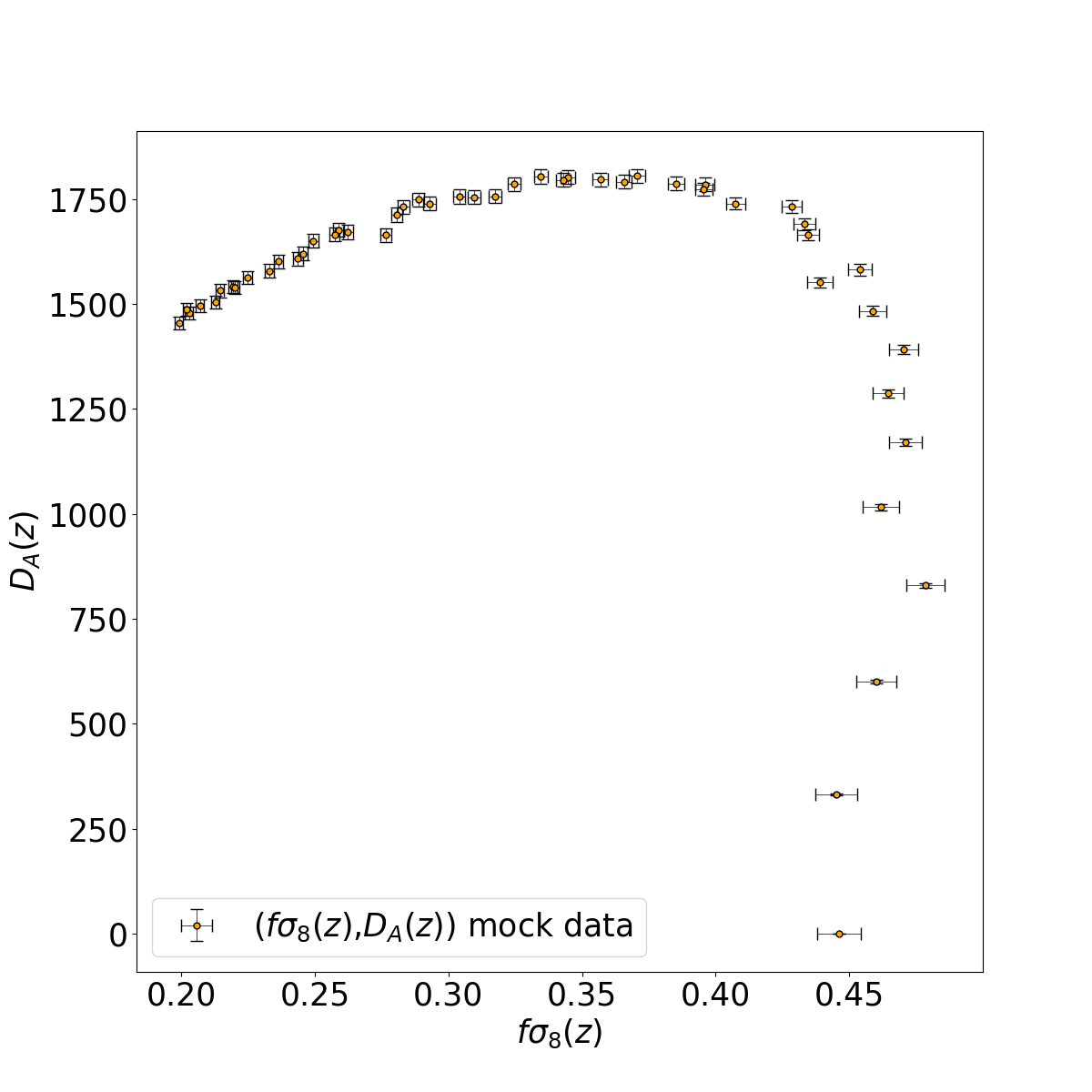}
	}
	\subfigure[]{
		\includegraphics[width=0.3\linewidth]{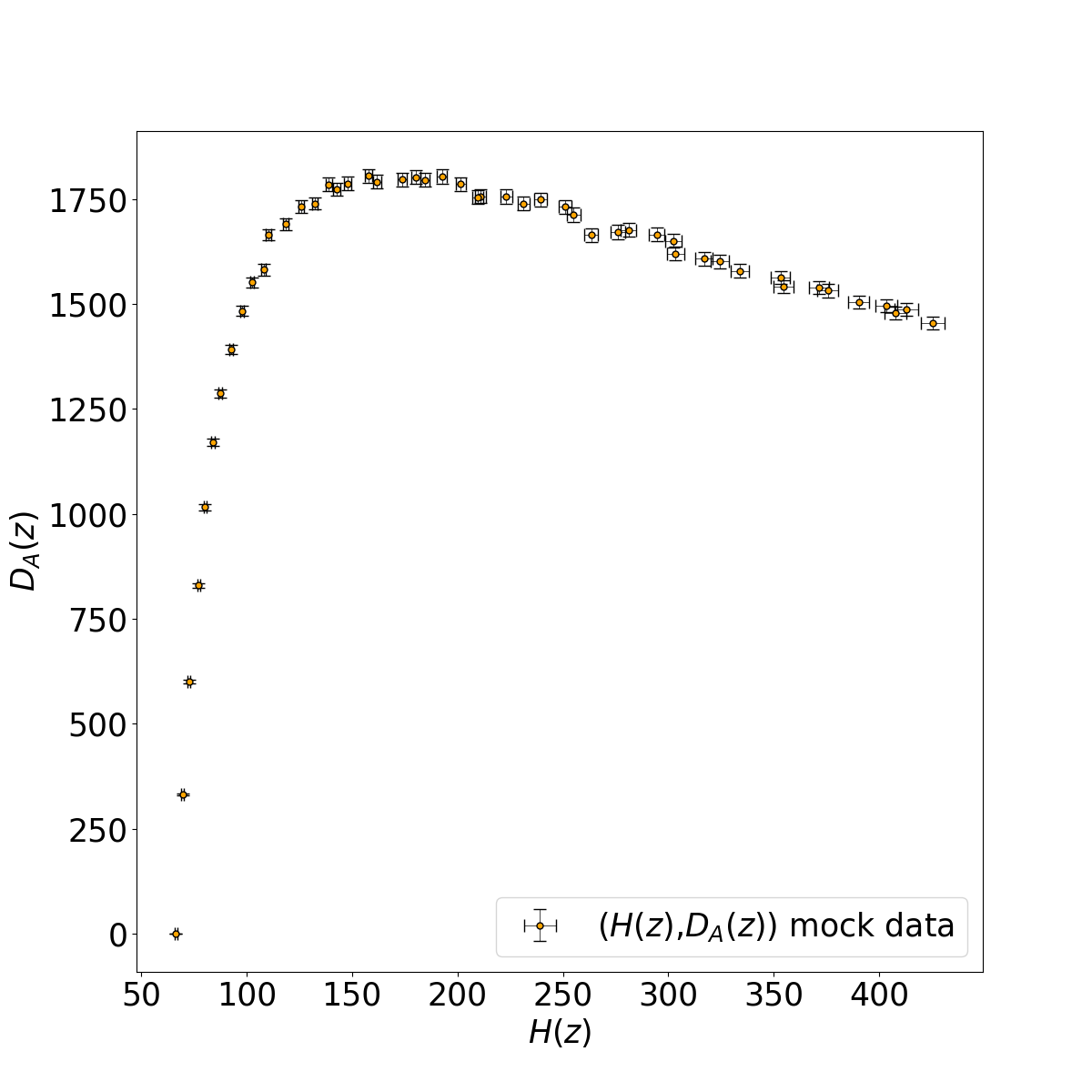}
	}
	\caption{The Gaussian sampling results of (a) Hubble constant $H_{0}$, (b) matter density parameter $\Omega_{m0}$ as well as (c) matter fluctuation amplitude $\sigma_{8}$ and the joint mock datasets of (d) $\left(H(z),f\sigma_{8}(z) \right)$, (e) $\left(f\sigma_{8}(z),D_{A}(z)\right)$, and (f) $\left(H(z),D_{A}(z)\right)$.}
	\label{fig:3-before}
\end{figure*}

\begin{figure}[ht!]
	\centering
	\includegraphics[width=1\linewidth]{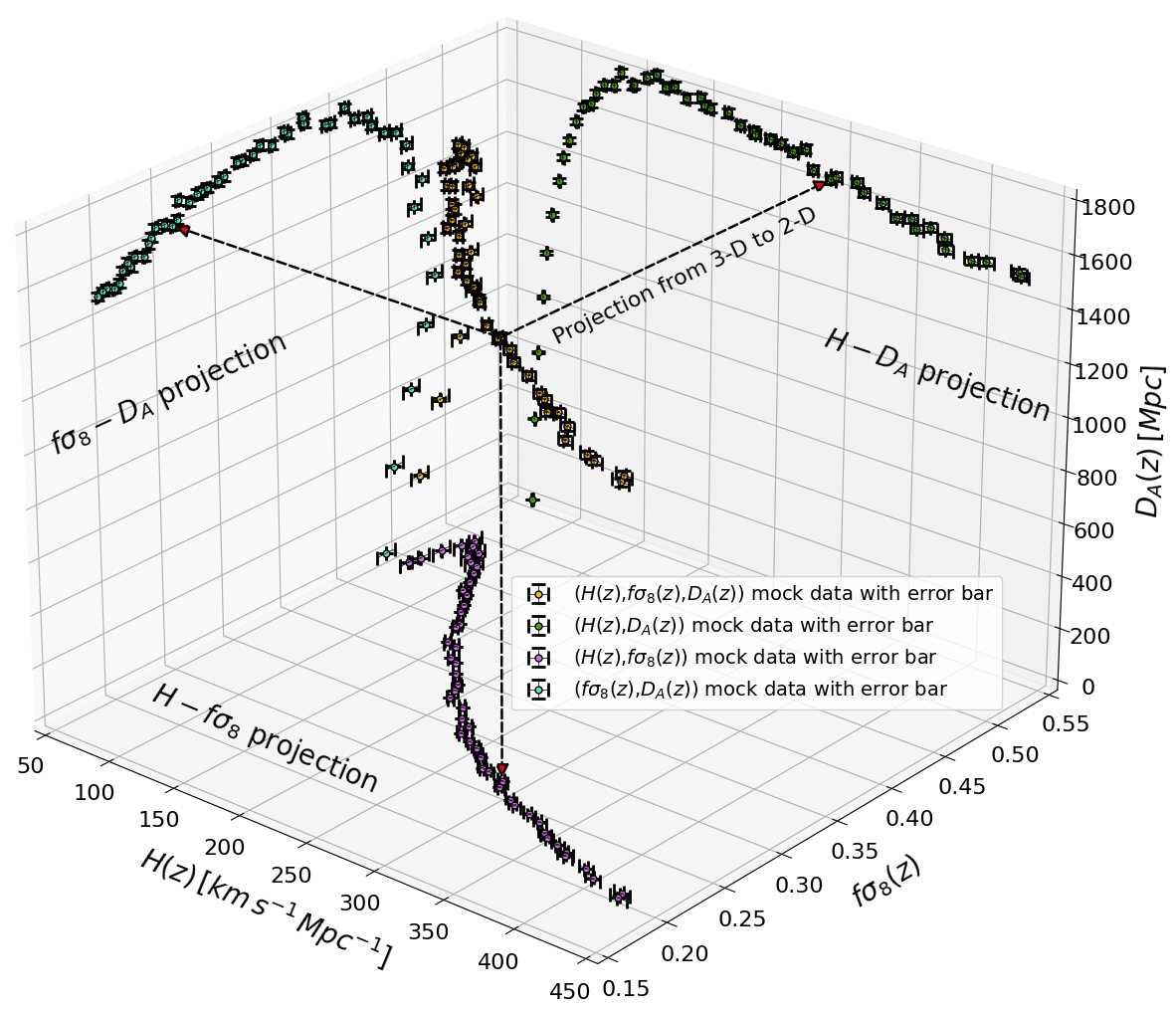}
	\caption{Observable curves in 3D $\left(H,f\sigma_{8},D_{A}\right)$ space and 2D projections respectively. Data points and errors are simulated based on a fiducial flat $\Lambda$CDM model assuming the Planck 2018 cosmological parameters.}
	\label{fig:3}
\end{figure}

To validate the efficacy of our joint constraint method and assess its capacity to constrain parameters of the cosmological model, we apply it to a set of carefully simulated data. Whereupon, we generate the mock data $\mathcal{F}\left(z ; \boldsymbol{\theta}\right)$ with inferred parameter values from Planck 2018 results: Hubble constant $H_0=67.4\pm 0.5 \mathrm{~km} \mathrm{~s}^{-1} \mathrm{~Mpc}^{-1}$, matter density parameter $\Omega_{\mathrm{m}}=0.315 \pm 0.007$ and matter fluctuation amplitude $\sigma_8=0.811 \pm 0.006$ in the flat $\Lambda$CDM model as the cosmic curvature $\Omega_K=0.001 \pm 0.002$ \citep{2020A&A...641A...6P}. These parameters are believed to conform to the Gaussian distribution. After making Gaussian sampling, sufficient sampling results, which are to ensure that the mean and variance of the sample from the sampling are as consistent as possible with the Gaussian distribution, are randomly selected and substituted into theoretical $\mathcal{F}\left(z ; \boldsymbol{\theta}\right)$ at different redshifts $z$, and then we calculate their mean and variance, which correspond to the truth value and variance of mock data at different redshifts. We show our sampling and mock results in Fig. \ref{fig:3-before}. The upper panel: the $H(z)$ uses the parameter set $\left\lbrace H_0, \Omega_{m0}\right\rbrace$, the $f\sigma_{8}(z)$ uses the parameter set $\left\lbrace \Omega_{m0},\sigma_{8}\right\rbrace$ and the $D_{A}(z)$ uses the parameter set $\left\lbrace H_0, \Omega_{m0}\right\rbrace$. And for the lower panel: the mock datasets are obtained from the Gaussian sampling results at each redshift $z$ which can be judged to conform to the Gaussian distribution using the $\chi^2$ goodness of fit with the rejection region $W=\left\lbrace \chi^2>\chi^2_{0.95} \right\rbrace$. The three figures represent the projection of our three-dimensional observables space in different directions.
And, we illustrate them here in three dimensions and project them onto the corresponding coordinate plane in Fig. \ref{fig:3}. This figure also illustrates our processing ideas for various types of observation data: first, a high-dimensional curve with an error bar is formed and then projected onto the corresponding two-dimensional plane for easy processing.

Subsequently, considering that some current observations cannot measure the redshift and no observation can observe all the observed space $V_{\mathrm{obs}}^{\mathrm{joint}}$ for a certain object at the same time, we regard the reconstructed results of different observables and data at different redshifts in the same range as the ``real" observational data. We consider the total observation atlas to be the most informative $H(\mathcal{F}_{obs})=-\sum_{i=1}^N p\left(\mathcal{F}_{i,obs}\right) \log p\left(\mathcal{F}_{i,obs}\right)$ \citep{1949mtc..book.....S,1950PhT.....3i..31S} and it is not possible to increase the quantity of information we can use through specific selection and combination of observational data. Predictions beyond present observational data are also inaccurate because we can not know what the observational results of the next observation will be. 

\begin{figure}[ht!]
	\centering
	\includegraphics[width=1\linewidth]{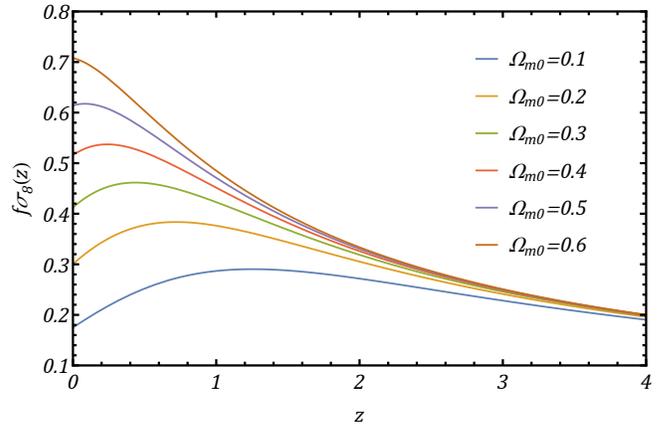}
	\caption{The theoretical curves of $f\sigma_{8}(z)$ vary with redshift $z$ from 0 to 4, which are colored corresponding to $\Omega_{m0}=0.1,0.2,0.3,0.4,0.5,0.6$ respectively.}
	\label{fig:apthree-1-1}
\end{figure}

\begin{figure}[ht!]
	\centering
	\includegraphics[width=1\linewidth]{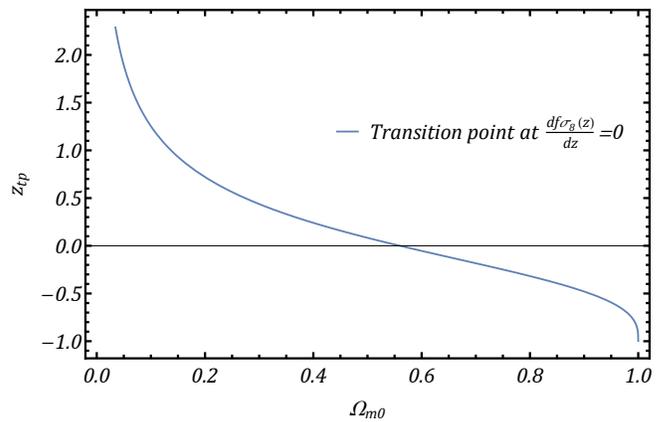}
	\caption{The transition point curve of $f\sigma_{8}$ as a function of matter density parameter $\Omega_{m0}$ from 0 to 1.}
	\label{fig:apthree-1-2}
\end{figure}

\begin{figure*}[ht!]
	\centering
	\subfigure[]{
		\includegraphics[width=0.33\linewidth]{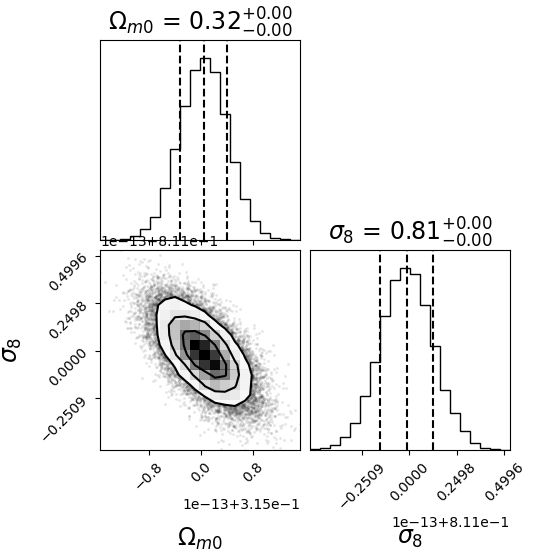}
	}
	\subfigure[]{
		\includegraphics[width=0.33\linewidth]{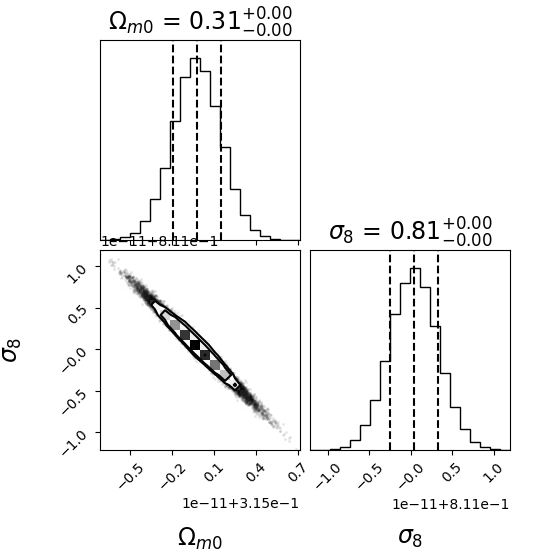}
	}
	\subfigure[]{
		\includegraphics[width=0.33\linewidth]{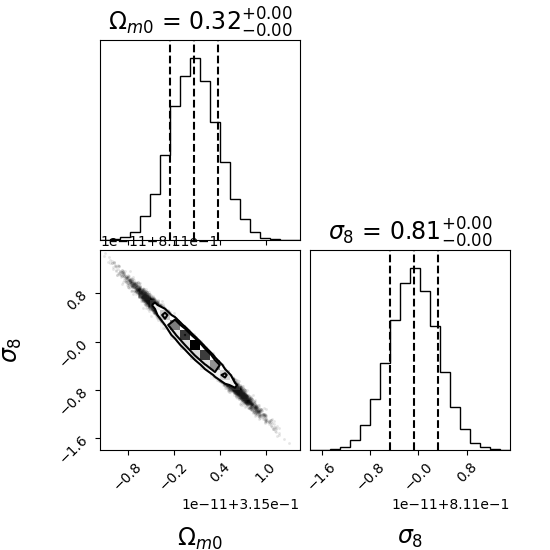}
	}
	\subfigure[]{
		\includegraphics[width=0.33\linewidth]{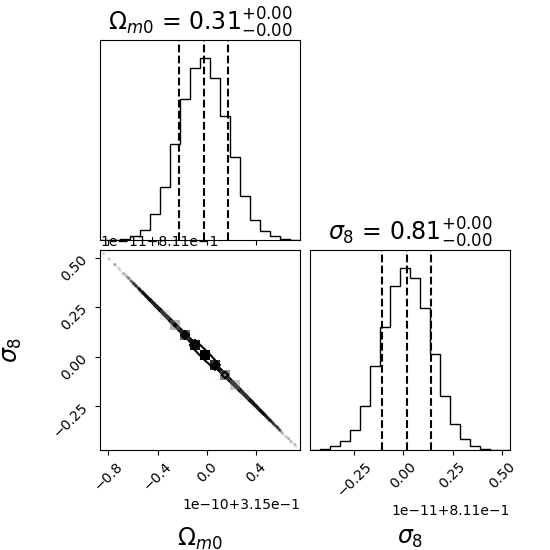}
	}
	\caption{The $68\%$, $95\%$, $99\%$ confidence regions of the joint and marginal posterior probability distributions of $\Omega_{m0}$ and $\sigma_{8}$ that is estimated from parameter constraint with (a) the total mock data $z\in[0,4]$, (b) the part mock data ordering before the before the transition point $z\in[0,z_{ep}]$, (c) the part mock data containing the transition point $z\in[0.2,0.6]$, as well as (d) the part mock data ordering after the transition point $z\in[3.6,4]$.}
	\label{fig:apthree-2}
\end{figure*}

Besides, the different segments of the same observational dataset cause the the difference in the results of the parameter constraints. This is a pragmatic issue. Occasionally, our observations can not completely provide enough observational data to cover the whole theoretical curve and only give partial redshift interval results. For a non-monotonic theoretical curve such as $f\sigma_{8}(z)$, the transition point of its monotonic variation is very sensitive to constraining cosmological parameters, so the observation near the transition point is very important. We illustrate the theoretical curve of $f\sigma_{8}(z)$ in Fig. \ref{fig:apthree-1-1} and transition point curve of $f\sigma_{8}(z)$ in Fig. \ref{fig:apthree-1-2} by setting the parameters $\lambda=1.3$, $\sigma_{8}=0.811$, $\gamma=0.78$ and $\beta=1.03$. Then, consider an extreme case where we use the parameter settings $\lambda=1.3$, $\sigma_{8}=0.811$, $\gamma=0.78$, $\beta=1.03$ and $\Omega_{m0}=0.315$ here, but set the observational error to zero and generate new mock data with redshift $z\in[0,4]$, which we name this dataset extreme-data to distinguish it from the previous mock data. In this manner, when we use the extreme-data to constrain cosmological parameters, we are not affected by error variations. The results of the parameter constraints are shown in Fig. \ref{fig:apthree-2}. We can find that there are minor differences between the cosmological parameter constraints of the complete extreme-data set and those of the piecewise extreme-data set, but the parameters exhibit clear degeneracies. After considering the piecewise extreme-data with the transition point, the parameter degeneracies are attenuated compared with other segments. So we should have an observation that can obtain as much cosmological information about the same object as possible to cancel out the effect caused by the different redshift intervals, such as the transition points of $f\sigma_{8}(z)$ and $D_{A}$.

\begin{figure*}[ht!]
	\centering
	\subfigure[]{
		\includegraphics[width=0.8\linewidth]{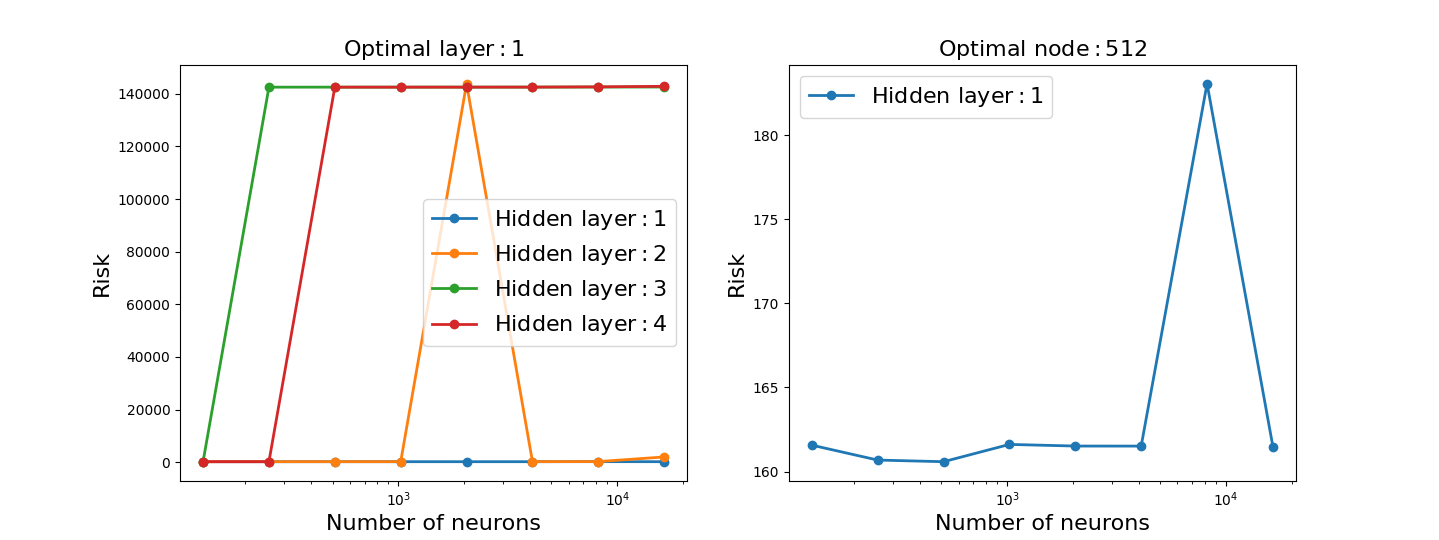}
	}
	\subfigure[]{
		\includegraphics[width=0.8\linewidth]{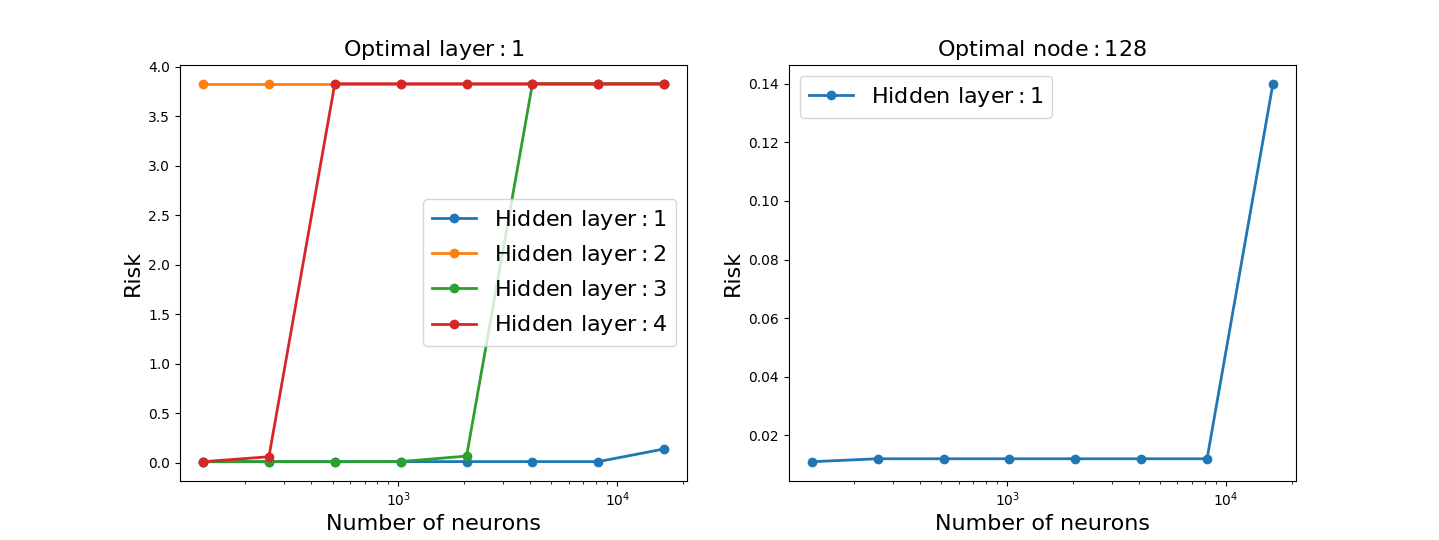}
	}
	\subfigure[]{
		\includegraphics[width=0.8\linewidth]{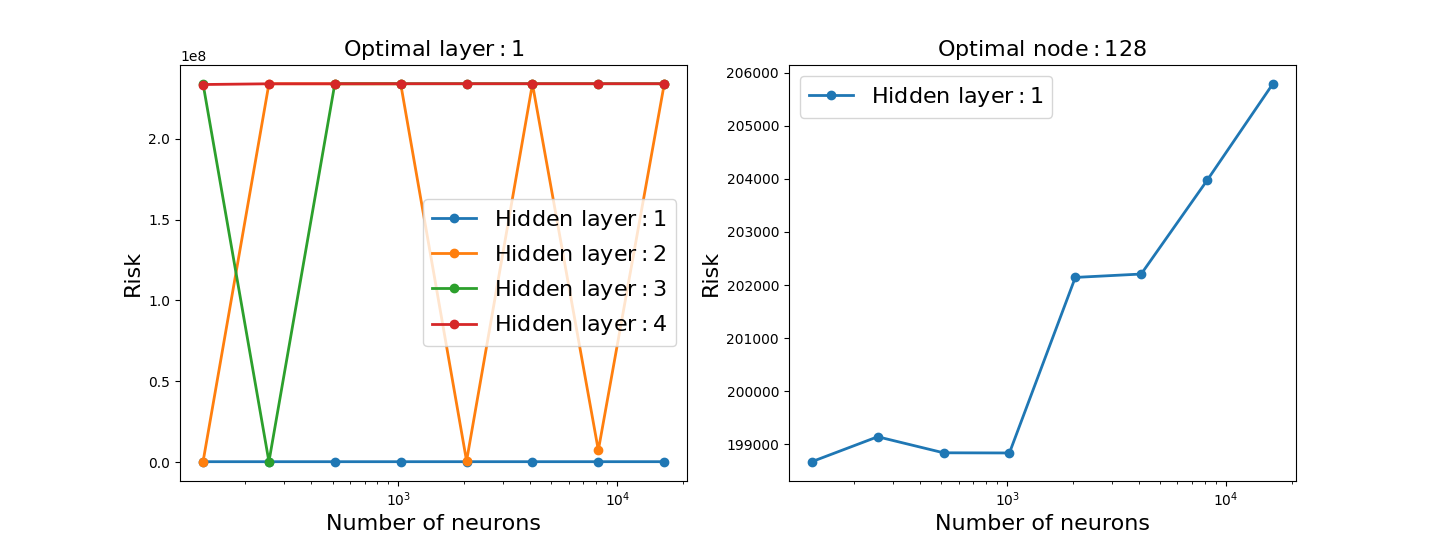}
	}
	\caption{The optimized hyperparameters of a network in reconstructing observable with the number of hidden layers $N_{hl}$ and the number of neurons $N_{ne}$ for (a) Hubble expansion rate $H(z)$, $N_{hl}=1$, $N_{ne}=512$, (b) growth history of the universe $f\sigma_{8}(z)$, $N_{hl}=1$, $N_{ne}=128$, as well as (c) the angular diameter distance $D_{A}(z)$, $N_{hl}=1$, $N_{ne}=128$.}
	\label{fig:aptwo-1}
\end{figure*}

\begin{figure*}[ht!]
	\centering
	\subfigure[]{
		\includegraphics[width=0.3\linewidth]{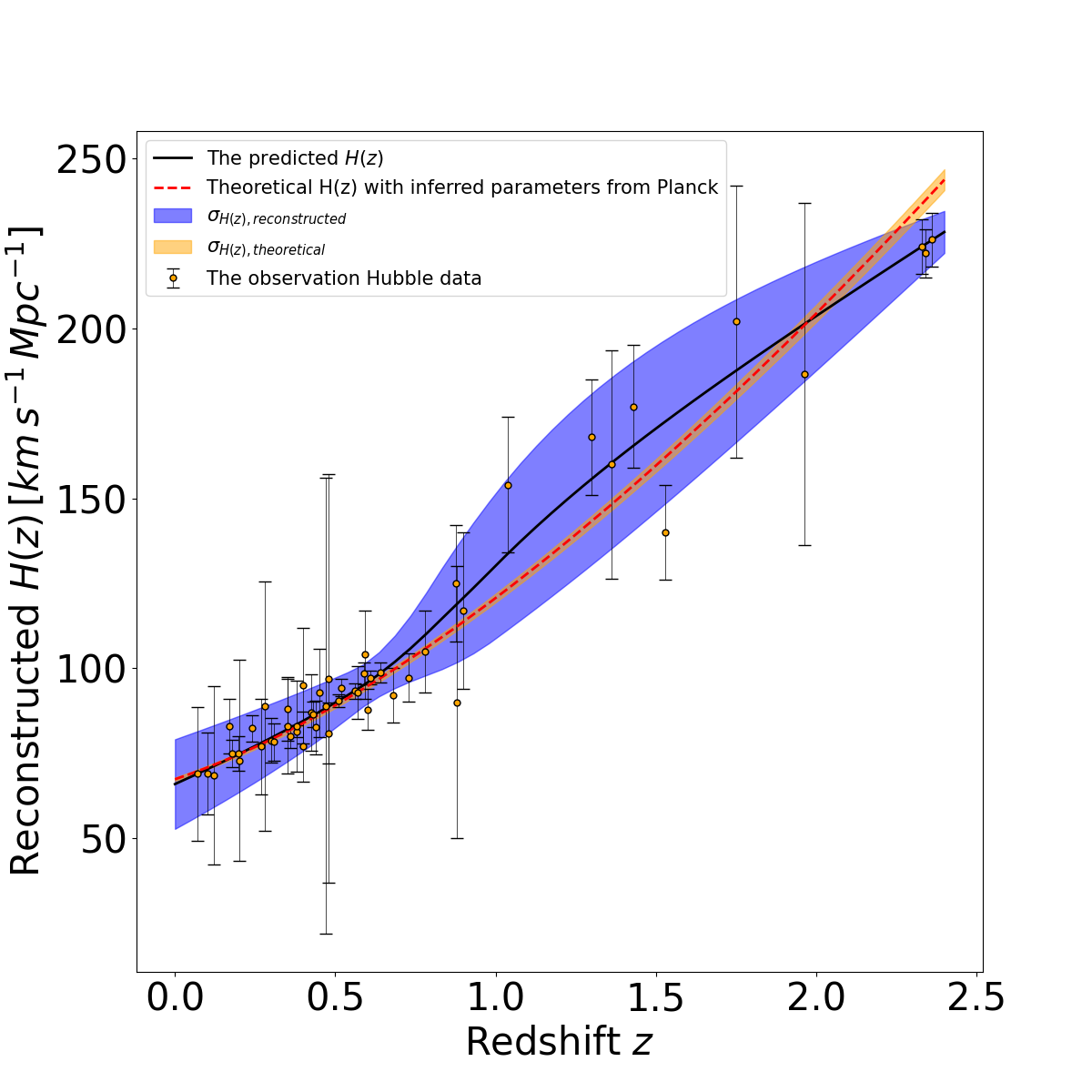}
	}
	\subfigure[]{
		\includegraphics[width=0.3\linewidth]{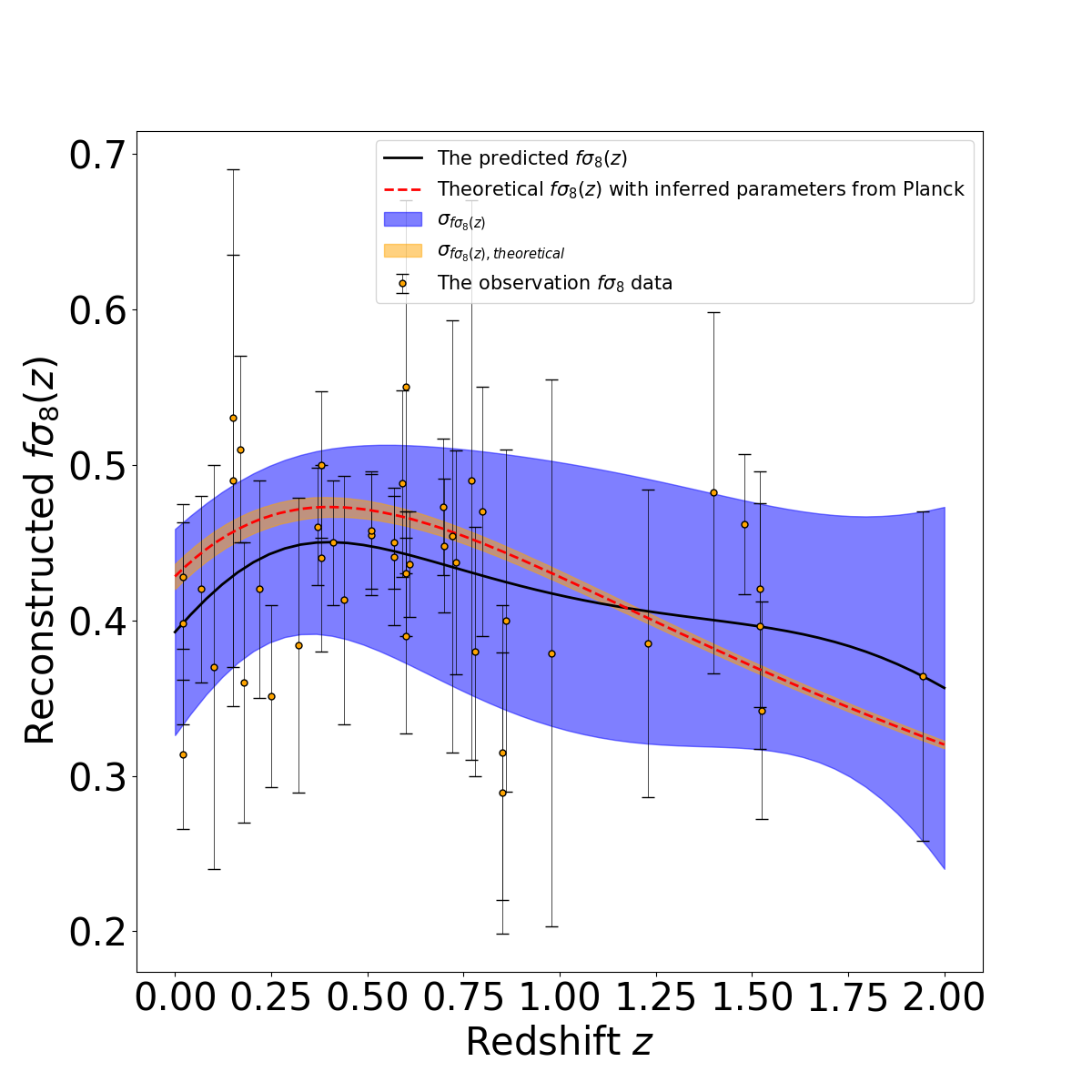}
	}
	\subfigure[]{
		\includegraphics[width=0.3\linewidth]{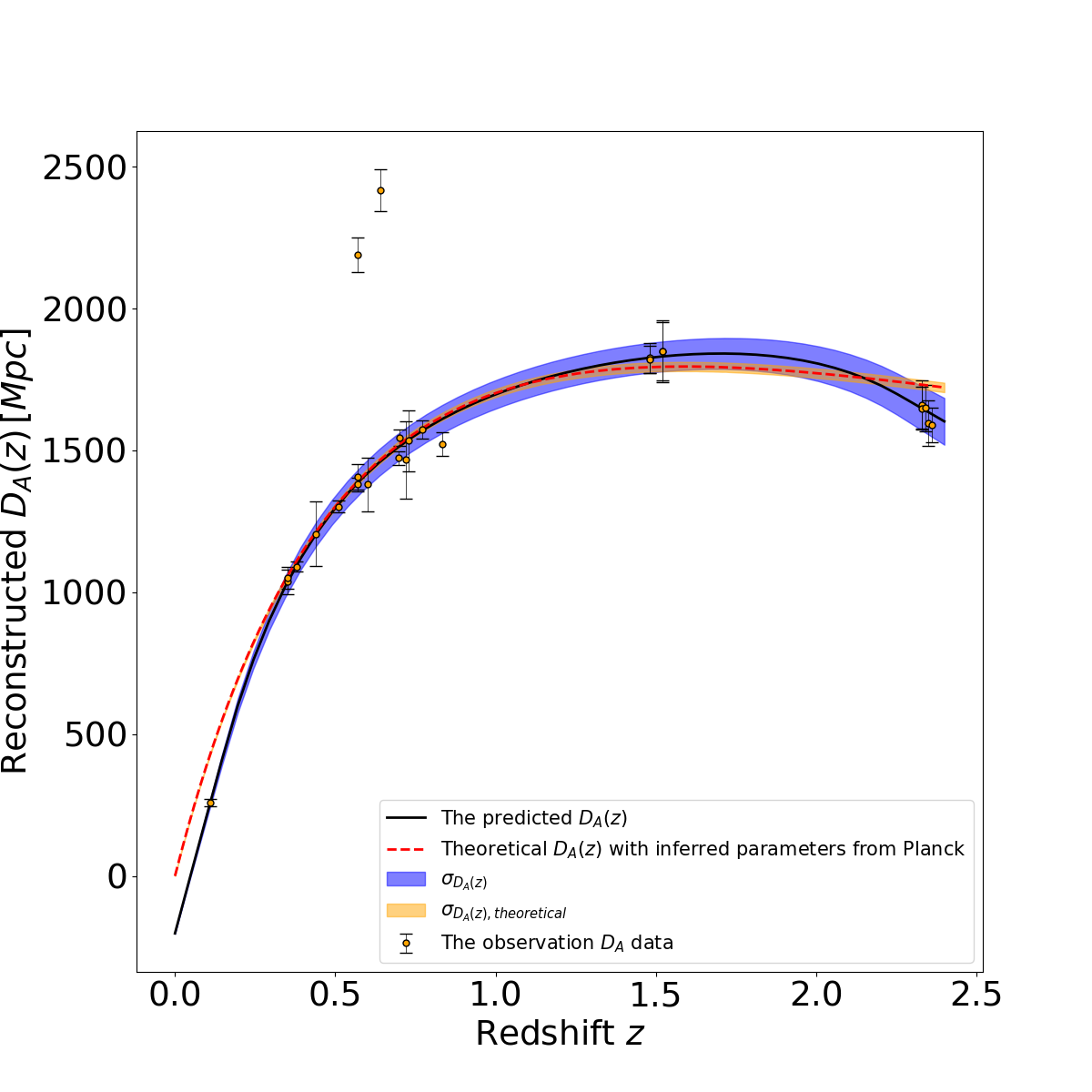}
	}
	\caption{The reconstructed functions of (a) $H(z)$, (b) $f\sigma_{8}(z)$, and (c) $D_{A}(z)$ labelled by the black solid lines and the corresponding $1\sigma$ error with neural networks that have one hidden layer shown by the blue error areas. The yellow dots with black error bars represent the observational data, the red dashed lines with orange error areas correspond to the fiducial flat $\Lambda$CDM models $\left\lbrace H(z),f\sigma_{8}(z),D_{A}(z)\right\rbrace$ with parameters $\left\lbrace H_{0},\Omega_{m0},\sigma_{8}\right\rbrace$ inferred from Planck 2018 results respectively.}
	\label{fig:4}
\end{figure*}

After evaluating the aforementioned factors, we choose $z\in[0.11,1.944]$ as the redshift range and we have the sufficient observational data points containing the transition points. We use a nonparametric approach for reconstructing functions from observational data using an Artificial Neural Network (ANN) called Reconstruct Functions with ANN (ReFANN) \citep{2020ApJS..246...13W}. One can estimate the optimal hyperparameters selected before using ANN for the reconstruction of functions, such as the optimal number of hidden layers and the number of neurons in the hidden layer, based on the minimize the risk \citep{2001astro.ph.12050W} or the Bayesian optimization \citep{2012arXiv1206.2944S}. The observational datasets of $H(z)$, $f\sigma_{8}(z)$ and $D_{A}(z)$ are enumerated in Table. \ref{table:I}, Table. \ref{table:II} and Table. \ref{table:III}, respectively. It is worth noting that the three datasets are independent of each other, the ratio of data volume between datasets is constant, and the three datasets are equally weighted. For example, the ratio of data volume between $H(z)$ and $f\sigma_{8}(z)$ is $52/45$, which is a constant close to 1, meaning the volume of data is of the same order of magnitude. There will not be a situation where one dataset overwrites another dataset, which will not affect the correlation of parameter inference. We conduct hyperparameter tuning for different observables with mock datasets respectively to find the optimal number of hidden layers and the number of neurons in the hidden layer, and our results are shown in Fig. \ref{fig:aptwo-1}. In Fig. \ref{fig:aptwo-1}, we set the initial learning rate is $0.01$ and it decreases with the number of iterations set to $30000$ which is large enough to assure the loss function on longer decreases. At the same time, we use the Exponential Linear Unit (ELU) as the activation function to optimize $H(z)$ and $D_{A}(z)$ and use the LogSigmoid as the activation function to optimize $f\sigma_{8}(z)$. Furthermore, all the activation functions contain no batch normalization and the batch size is specified at half the number of the mock data. In Fig. \ref{fig:aptwo-1} for the left panel: the optimal number of hidden layers is obtained from the average risk of the model for network structure. And for the right panel: the optimal number of neurons in the network with an optimized hidden layer. Using the optimized neural network, we reconstruct the three observables $H(z)$, $f\sigma_{8}(z)$ and $D_{A}(z)$ in redshift $z\in[0.11,1.944]$. We show our results of reconstructed functions in Fig. \ref{fig:4}.
\begin{table}[!ht]
	\centering
	\caption{The compiled independent $D_{A}(z)$ dataset.}
		\begin{tabular}{ccc}
			\hline \hline Redshift $z$ & $D_{A}(z)^{*} \pm 1 \sigma$ error & References \\
			\hline $0.11$ & $258.31 \pm 13.71$ &\cite{deCarvalho:2021azj} \\
			$0.35$ & $1037 \pm 44$ &\cite{Hemantha:2013sea} \\
			$0.35$ & $1050 \pm 38$ & \cite{2013MNRAS.431.2834X} \\
			$0.38$ & $1090.90 \pm 18.13$ & \cite{2021PhRvD.103h3533A} \\
			$0.44$ & $1205 \pm 114$ &\cite{2012MNRAS.425..405B} \\
			$0.51$ & $1302.02 \pm 20.47$ &\cite{2021PhRvD.103h3533A} \\
			$0.57$ & $1408 \pm 45$ & \cite{2014MNRAS.439.3504S} \\
			$0.57$ & $2190 \pm 61$ &\cite{2012MNRAS.426.2719R} \\
			$0.57$ & $1380 \pm 23$ & \cite{2014MNRAS.439.3504S} \\
			$0.6$ & $1380 \pm 95$ &\cite{2012MNRAS.425..405B} \\
			$0.64$ & $2418 \pm 73$ &\cite{2018MNRAS.477.5090P} \\
			$0.698$ & $1473.23 \pm 25.04$ &\cite{2021MNRAS.500..736B} \\
			$0.7$ & $1546.05 \pm 28.57$ & \cite{2021PhRvD.103h3533A} \\
			$0.72$ & $1466.5 \pm 136.6$ &\cite{2020MNRAS.492.4189I} \\
			$0.73$ & $1534 \pm 107$ &\cite{2012MNRAS.425..405B} \\
			$0.77$ & $1573.39 \pm 31.72$ &\cite{2020MNRAS.498.3470W} \\
			$0.835$ & $1521.85 \pm 41.02$ &\cite{2022PhRvD.105d3512A} \\
			$1.48$ & $1826.50 \pm 52.42$ & \cite{2021MNRAS.500.1201H} \\
			$1.48$ & $1821.11 \pm 47.47$ &\cite{2021PhRvD.103h3533A} \\
			$1.52$ & $1850.0 \pm 102.5$ &\cite{2018MNRAS.477.1639Z} \\
			$1.52$ & $1850 \pm 110$ &\cite{2018MNRAS.477.1604G} \\
			$2.33$ & $1661.63 \pm 83.97$ &\cite{2021PhRvD.103h3533A} \\
			$2.33$ & $1648.37 \pm 75.13$ &\cite{2021PhRvD.103h3533A} \\
			$2.34$ & $1650.18 \pm 82.05$ &\cite{deSainteAgathe:2019voe} \\
			$2.35$ & $1596.44 \pm 79.16$ &\cite{2019AA...629A..86B} \\
			$2.36$ & $1590 \pm 60$ &\cite{2014JCAP...05..027F} \\
			\hline
		\end{tabular}
	\begin{flushleft}
\tablecomments{
	* $D_{A}(z)$ are in the unit of $\mathrm{Mpc}$.}
	\end{flushleft}
	\label{table:I}
\end{table}

\begin{table}[]
	\centering
	\caption{The compiled independent $f\sigma_{8}(z)$ dataset.}
	\begin{tabular}{ccc}
		\hline \hline $z$ & $f \sigma_8(z) $  & References \\
		\hline $0.067$ & $0.420\pm 0.060$ & \cite{Beutler:2012px} \\
		$0.02$ & $0.428 \pm 0.0465$ &\cite{2017JCAP...05..015H} \\
		$0.02$ & $0.398 \pm 0.065$ & \cite{2012MNRAS.420..447T} \\
		$0.02$ & $0.314 \pm 0.048$ &\cite{2012ApJ...751L..30H} \\
		$0.10$ & $0.370 \pm 0.130$ &\cite{Feix:2015dla} \\
		$0.15$ & $0.490 \pm 0.145$ &\cite{Howlett:2014opa} \\
		$0.15$ & $0.530 \pm 0.160$ &\cite{2021PhRvD.103h3533A} \\
		$0.17$ & $0.510 \pm 0.060$ &\cite{2009JCAP...10..004S} \\
		$0.18$ & $0.360 \pm 0.090$ & \cite{Blake:2013nif} \\
		$0.22$ & $0.420\pm 0.070$ & \cite{2011MNRAS.415.2876B} \\
		$0.25$ & $0.3512 \pm 0.0583$ &\cite{Samushia:2011cs} \\
		$0.32$ & $0.384 \pm 0.095$ &  \cite{Sanchez:2013tga} \\
		$0.37$ & $0.4602 \pm 0.0378$ & \cite{Samushia:2011cs} \\
		$0.38$ & $0.440 \pm 0.060$ &  \cite{Blake:2013nif} \\
		$0.38$ & $0.500\pm 0.047$ &\cite{2021PhRvD.103h3533A} \\
		$0.41$ & $0.450\pm 0.040$ & \cite{2011MNRAS.415.2876B} \\
		$0.44$ & $0.413 \pm 0.080$ &\cite{2012MNRAS.425..405B} \\
		$0.51$ & $0.455\pm 0.039$ &\cite{2021PhRvD.103h3533A} \\
		$0.51$ & $0.458\pm 0.038$ & \cite{2017MNRAS.470.2617A} \\
		$0.57$ & $0.441\pm 0.044$ & \cite{2014MNRAS.439.3504S} \\
		$0.57$ & $0.447\pm 0.028$ &\cite{2014MNRAS.439.3504S} \\
		$0.59$ & $0.488 \pm 0.060$ &\cite{2016MNRAS.461.3781C} \\
		$0.60$ & $0.390 \pm 0.063$ &\cite{2012MNRAS.425..405B} \\
		$0.60$ & $0.550 \pm 0.120$ &\cite{2017AA...604A..33P} \\
		$0.60$ & $0.430\pm 0.040$ &\cite{2011MNRAS.415.2876B} \\
		$0.61$ & $0.436\pm 0.034$ &\cite{2017MNRAS.470.2617A} \\
		$0.698$ & $0.473\pm 0.044$ &\cite{2021MNRAS.500..736B} \\
		$0.70$ & $0.448\pm 0.043$ & \cite{2021PhRvD.103h3533A} \\
		$0.72$ & $0.454 \pm 0.139$ & \cite{2020MNRAS.492.4189I} \\
		$0.73$ & $0.437 \pm 0.072$ & \cite{2012MNRAS.425..405B} \\
		$0.77$ & $0.490\pm 0.180$ & \cite{2008Natur.451..541G} \\
		$0.78$ & $0.380\pm 0.080$ & \cite{2011MNRAS.415.2876B} \\
		$0.80$ & $0.470\pm 0.080$ & \cite{2017AA...601A.144R} \\
		$0.85$ & $0.315\pm 0.095$ &\cite{2021PhRvD.103h3533A} \\
		$0.85$ & $0.289\pm 0.091$ & \cite{2021MNRAS.501.5616D} \\
		$0.86$ & $0.400 \pm 0.110$ & \cite{2017AA...604A..33P} \\
		$0.978$ & $0.379 \pm 0.176$ & \cite{2019MNRAS.482.3497Z} \\
		$1.23$ & $0.385 \pm 0.099$ &\cite{2019MNRAS.482.3497Z} \\
		$1.40$ & $0.482 \pm 0.116$ & \cite{2016PASJ...68...38O} \\
		$1.48$ & $0.462\pm 0.045$ & \cite{2021PhRvD.103h3533A} \\
		$1.48$ & $0.476\pm 0.047$ & \cite{2020MNRAS.499..210N} \\
		$1.52$ & $0.420 \pm 0.076$ & \cite{2018MNRAS.477.1604G} \\
		$1.52$ & $0.396 \pm 0.079$ & \cite{2018MNRAS.480.2521H} \\
		$1.526$ & $0.342 \pm 0.070$ & \cite{2019MNRAS.482.3497Z} \\
		$1.944$ & $0.364 \pm 0.106$ & \cite{2019MNRAS.482.3497Z} \\	
		\hline
	\end{tabular}
	\label{table:II}
\end{table}

\begin{table}[]
	\centering
	\caption{The compiled independent $H(z)$ dataset.}
	\begin{tabular}{ccc}
		\hline\hline  $z$ & $H(z)^{*}$  & References \\
		\hline $0.07$ & $69 \pm 19.6$ & \cite{Zhang:2012mp} \\
		$0.1$ & $69 \pm 12$ &\cite{2005PhRvD..71l3001S} \\
		$0.12$ & $68.6 \pm 26.2$ &\cite{Zhang:2012mp} \\
		$0.17$ & $83 \pm 8$ & \cite{2005PhRvD..71l3001S} \\
		$0.1791$ & $75 \pm 5$ & \cite{2012JCAP...08..006M} \\
		$0.1993$ & $75 \pm 5$ &  \cite{2012JCAP...08..006M} \\
		$0.2$ & $72.9 \pm 29.6$ &\cite{Zhang:2012mp} \\
		$0.24$ & $82.37 \pm 3.94$ & \cite{2009MNRAS.399.1663G} \\
		$0.27$ & $77 \pm 14$ &  \cite{2005PhRvD..71l3001S} \\
		$0.28$ & $88.8 \pm 36.6$ & \cite{Zhang:2012mp} \\
		$0.3$ & $78.83 \pm 6.58$ &\cite{2014MNRAS.439.2515O} \\
		$0.31$ & $78.39 \pm 5.46$ & \cite{2017MNRAS.469.3762W} \\
		$0.35$ & $88.10 \pm 9.45$ &\cite{2013MNRAS.435..255C} \\
		$0.3519$ & $83 \pm 14$ & \cite{2012JCAP...08..006M} \\
		$0.36$ & $79.93 \pm 3.39$ &\cite{2017MNRAS.469.3762W} \\
		$0.38$ & $81.5 \pm 1.8$ & \cite{2017MNRAS.470.2617A} \\
		$0.3802$ & $83 \pm 13.5$ & \cite{2016JCAP...05..014M} \\
		$0.4$ & $95 \pm 17$ & \cite{2005PhRvD..71l3001S} \\
		$0.4004$ & $77 \pm 10.2$ &  \cite{2016JCAP...05..014M} \\
		$0.4247$ & $87.1 \pm 11.2$ & \cite{2016JCAP...05..014M} \\
		$0.43$ & $86.45 \pm 3.68$ &\cite{2009MNRAS.399.1663G} \\
		$0.44$ & $82.6 \pm 7.8$ & \cite{2012MNRAS.425..405B} \\
		$0.4497$ & $92.8 \pm 12.9$ & \cite{2016JCAP...05..014M} \\
		$0.47$ & $89 \pm 67$ & \cite{2017MNRAS.467.3239R} \\
		$0.4783$ & $80.9 \pm 9$ & \cite{2016JCAP...05..014M} \\
		$0.48$ & $97 \pm 62$ &  \cite{2010ApJS..188..280S} \\
		$0.51$ & $90.4 \pm 1.9$ &\cite{2017MNRAS.470.2617A} \\
		$0.52$ & $94.35 \pm 2.65$ &  \cite{2017MNRAS.469.3762W} \\
		$0.56$ & $93.33 \pm 2.32$ & \cite{2017MNRAS.469.3762W} \\
		$0.57$ & $92.9 \pm 7.8$ & \cite{2014MNRAS.439...83A} \\
		$0.59$ & $98.48 \pm 3.19$ & \cite{2017MNRAS.469.3762W} \\
		$0.5929$ & $104 \pm 13$ & \cite{2012JCAP...08..006M} \\
		$0.6$ & $87.9 \pm 6.1$ &  \cite{2012MNRAS.425..405B} \\
		$0.61$ & $97.3 \pm 2.1$ &  \cite{2017MNRAS.470.2617A} \\
		$0.64$ & $98.82 \pm 2.99$ & \cite{2017MNRAS.469.3762W} \\
		$0.6797$ & $92 \pm 8$ & \cite{2012JCAP...08..006M} \\
		$0.73$ & $97.3 \pm 7$ & \cite{2012MNRAS.425..405B} \\
		$0.7812$ & $105 \pm 12$ &\cite{2012JCAP...08..006M} \\
		$0.8$ & $113.1 \pm 15.1$ &\cite{2023ApJS..265...48J}\\
		$0.8754$ & $125 \pm 17$ &\cite{2012JCAP...08..006M} \\
		$0.88$ & $90 \pm 40$ &  \cite{2010ApJS..188..280S} \\
		$0.9$ & $117 \pm 23$ &\cite{2005PhRvD..71l3001S} \\
		$1.037$ & $154 \pm 20$ &\cite{2012JCAP...08..006M} \\
		$1.3$ & $168 \pm 17$ & \cite{2005PhRvD..71l3001S} \\
		$1.363$ & $160 \pm 33.6$ & \cite{2015MNRAS.450L..16M} \\
		$1.43$ & $177 \pm 18$ & \cite{2005PhRvD..71l3001S} \\
		$1.53$ & $140 \pm 14$ &\cite{2005PhRvD..71l3001S} \\
		$1.75$ & $202 \pm 40$ &\cite{2005PhRvD..71l3001S} \\
		$1.965$ & $186.5 \pm 50.4$ & \cite{2015MNRAS.450L..16M} \\
		$2.33$ & $224 \pm 8$ & \cite{2017AA...603A..12B} \\
		$2.34$ & $222 \pm 7$ &\cite{2015AA...574A..59D} \\
		$2.36$ & $226 \pm 8$ & \cite{2014JCAP...05..027F} \\
		\hline
	\end{tabular}
	\tablecomments{
		* $H(z)$ in the unit of $\mathrm{km~s}^{-1}\;\mathrm{Mpc}^{-1}$.}
	\label{table:III}
\end{table}


\section{Result and Discussion}
\label{Rad}
\begin{figure*}[htbp]
	\centering
	\subfigure[]{
		\includegraphics[width=0.3\linewidth]{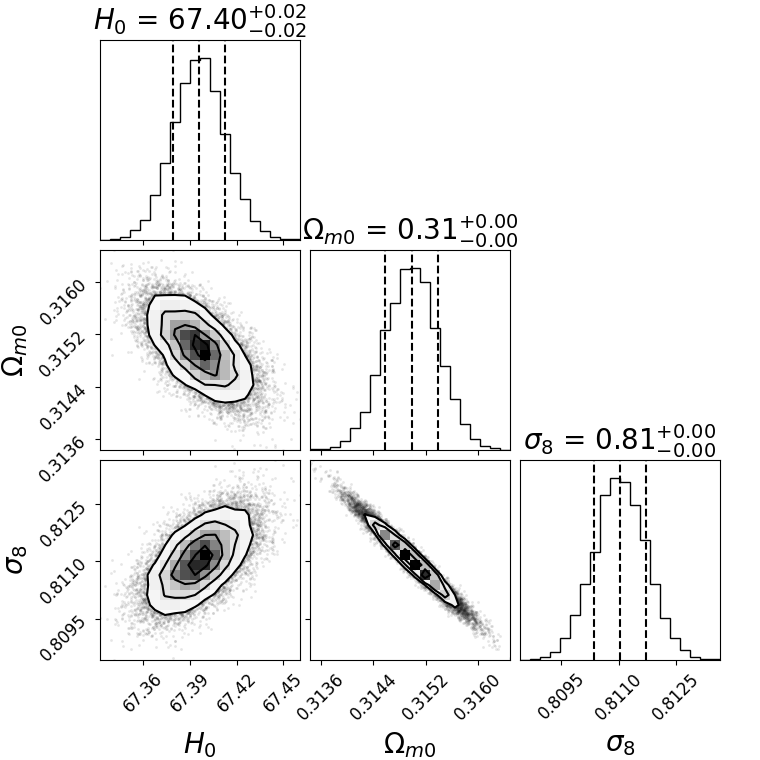}
	}
	\subfigure[]{
		\includegraphics[width=0.3\linewidth]{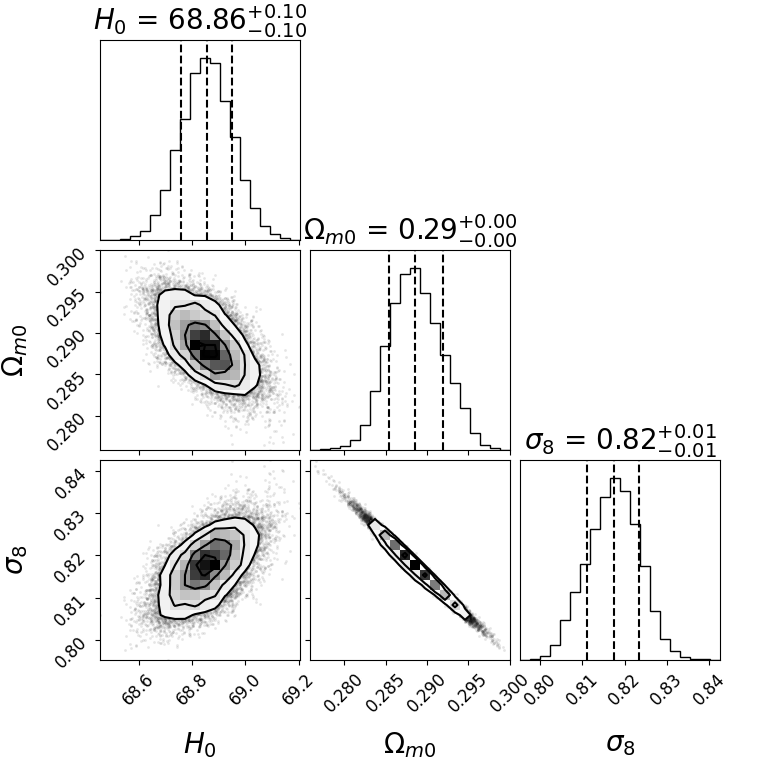}
	}
	\subfigure[]{
		\includegraphics[width=0.3\linewidth]{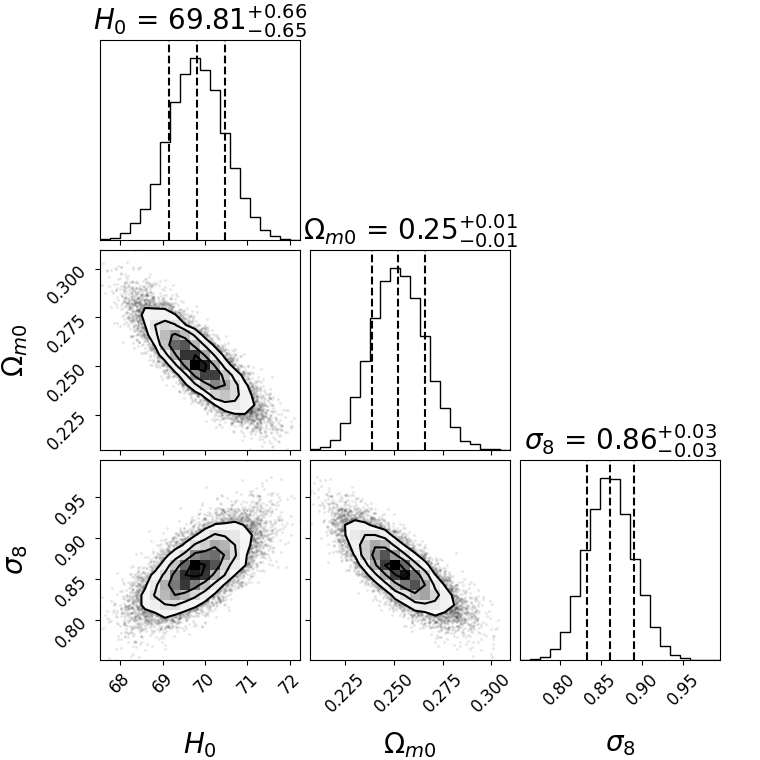}
	}
	\caption{The $68\%$, $95\%$, $99\%$ confidence regions of the joint and marginal posterior probability distributions of $H_0$, $\Omega_{m0}$ and $\sigma_{8}$ that is estimated from (a) our joint constraint method with the mock data, (b) observational data, as well as (c) the traditional combined method with observational data.}
	\label{fig:5}
\end{figure*}

In this section, we present the results obtained using our joint constraint method, as well as some discussion and extension of our method. We assume the priors $H_0\in[0,100]$, $\Omega_{m0}\in[0,1]$ and $\sigma_{8}\in[0,1]$ to constrain cosmological parameters with mock data and reconstructed data via our model $\mathcal{T}\left( V_{\mathrm{obs}}^{\mathrm{joint}}\right)$ in Markov chain Monte Carlo (MCMC) respectively. Here, we use the Python implementation of the affine-invariant ensemble sampler for Markov chain Monte Carlo (emcee) to obtain the estimated posterior \citep{2013PASP..125..306F}. The posteriors of mock data and reconstructed data are shown in Fig. \ref{fig:5}. With mock data using our joint constraint method, we find $H_0=67.40\pm0.02 \mathrm{~km} \mathrm{~s}^{-1} \mathrm{~Mpc}^{-1}$, $\Omega_{m0}=0.3150\pm0.0004$, $\sigma_{8}=0.8110\pm0.0007$. With observational data using our joint constraint method, we find $H_0=68.7\pm0.1\mathrm{~km} \mathrm{~s}^{-1}\mathrm{~Mpc}^{-1}$, $\Omega_{m0}=0.289\pm0.003$, $\sigma_{8}=0.82\pm0.01$. With observational data using the traditional combined method, we find $H_0=69.8\pm0.6\mathrm{~km} \mathrm{~s}^{-1} \mathrm{~Mpc}^{-1}$, $\Omega_{m0}=0.25\pm0.01$, $\sigma_{8}=0.86\pm0.03$.

\begin{figure*}[htbp]
	\centering
	\subfigure[]{
		\includegraphics[width=0.3\linewidth]{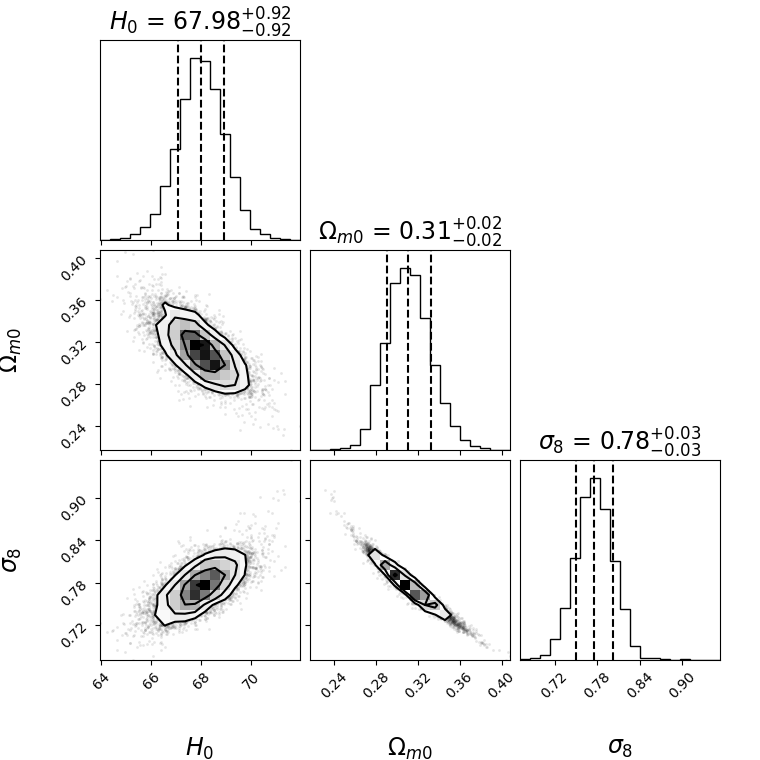}
	}
	\subfigure[]{
		\includegraphics[width=0.3\linewidth]{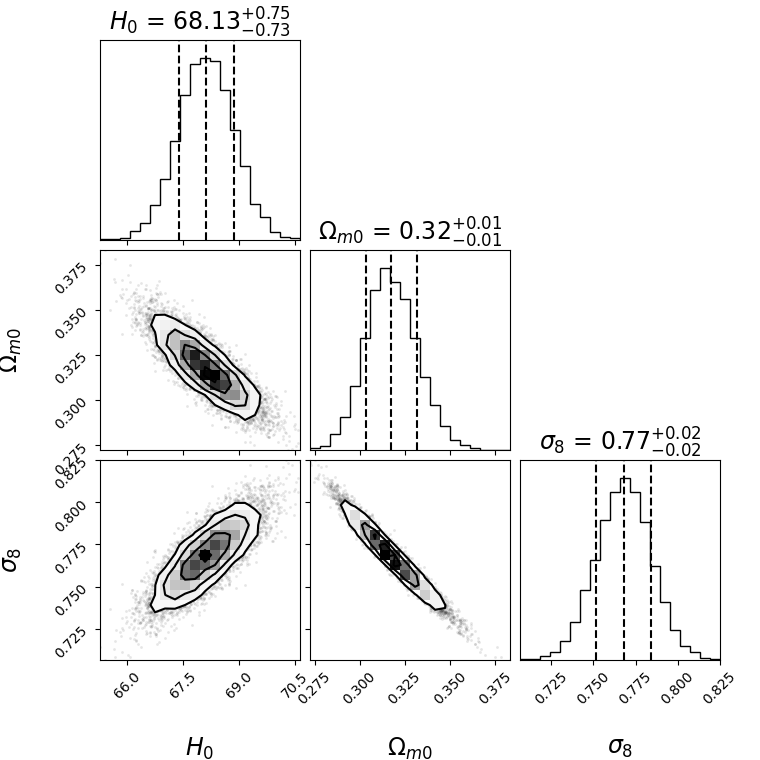}
	}
	\subfigure[]{
		\includegraphics[width=0.3\linewidth]{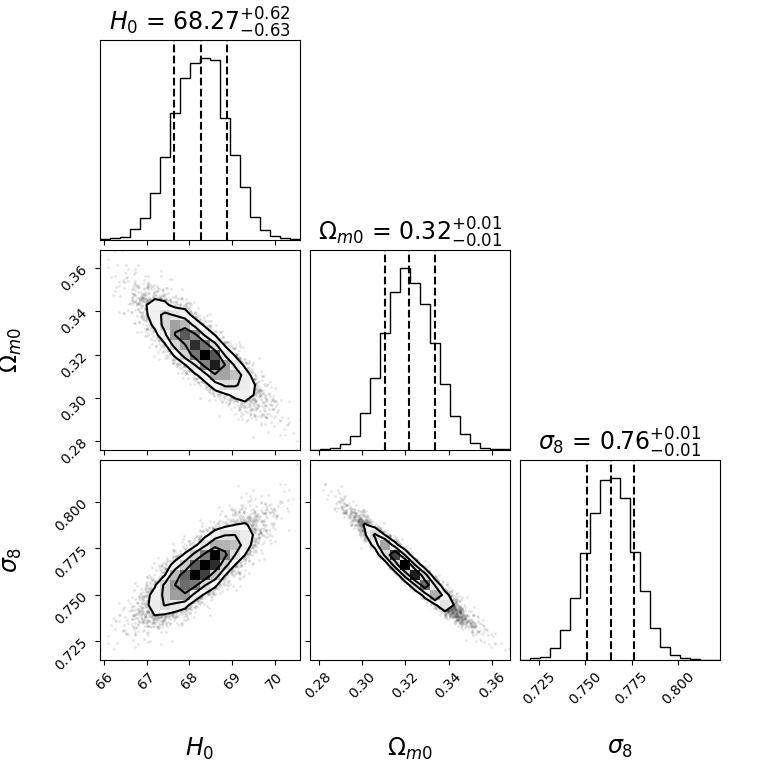}
	}
	\caption{The $68\%$, $95\%$, $99\%$ confidence regions of the joint and marginal posterior probability distributions of $H_0$, $\Omega_{m0}$ and $\sigma_{8}$ that is estimated from our joint constraint method (a) with 10 points in the observational data, (b) with 15 points in the observational data, as well as (c) with 20 points in the observational data.}
	\label{fig:sinp}
\end{figure*}

To our knowledge, the study reported here is the latest joint method to constrain cosmological parameters from multiple observables spanning in $M$-dimensional joint space. Since we make full use of the observational data, the errors of our inferential results are lower than that of traditional combined method and also break the parametric degeneracies to some extent. Moreover, our method requires fewer observational data when the results are nearly identical to those constrained by the traditional combined method because we project one observational point onto a pair of observational surface using Eq. (\ref{eq5}) $\mathcal{F}_{k,obs}\stackrel{\mathcal{T}}\mapsto\left(\mathcal{F}_{i,k,obs},\mathcal{F}_{j,k,obs} \right)$ which the observational information can be used more completely with observational error $\sigma_{k,obs}^2=\sigma_{i,k,obs}^{2}+\sigma_{j,k,obs}^{2}$ under Cartesian coordinate system. We show our results of this in Fig. \ref{fig:sinp}. With ten mock data points using our joint constraint method, we find $H_0=68.0 \pm 0.9\mathrm{~km} \mathrm{~s}^{-1} \mathrm{~Mpc}^{-1}$, $\Omega_{m0}=0.31 \pm 0.02$, $\sigma_{8}=0.78 \pm 0.03$. With fifteen observational data points, we find $H_0=68.1^{+0.8}_{-0.7} \mathrm{~km} \mathrm{~s}^{-1}\mathrm{~Mpc}^{-1}$, $\Omega_{m0}=0.32 \pm 0.01$, $\sigma_{8}=0.77 \pm 0.02$. With twenty observational data points, we find $H_0=68.3 \pm 0.6 \mathrm{~km} \mathrm{~s}^{-1} \mathrm{~Mpc}^{-1}$, $\Omega_{m0}=0.32 \pm 0.01$, $\sigma_{8}=0.76 \pm 0.01$.

\begin{figure*}[ht!]
	\centering
	\subfigure[]{
		\includegraphics[width=0.3\linewidth]{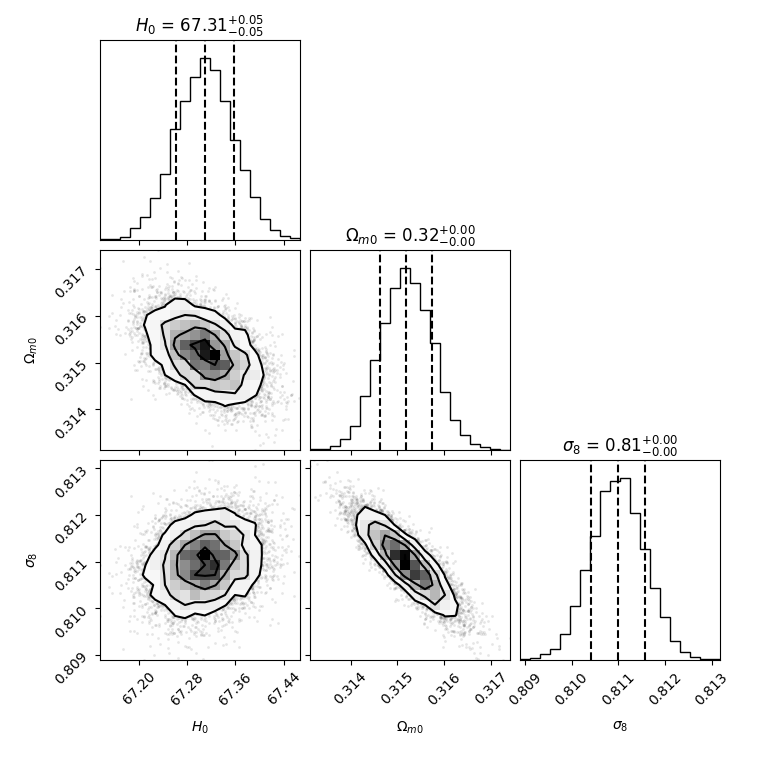}
	}
	\subfigure[]{
		\includegraphics[width=0.3\linewidth]{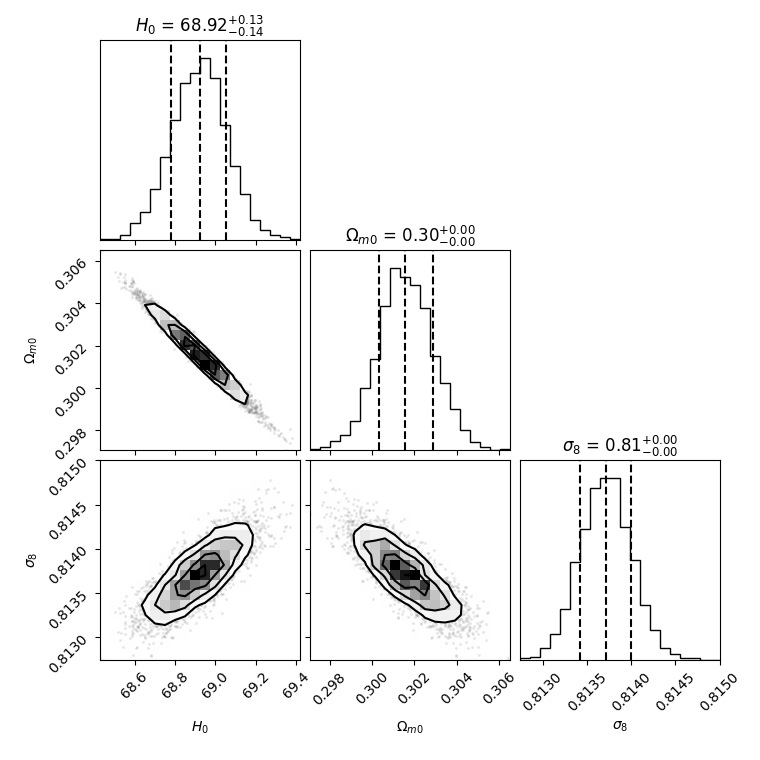}
	}
	\subfigure[]{
		\includegraphics[width=0.3\linewidth]{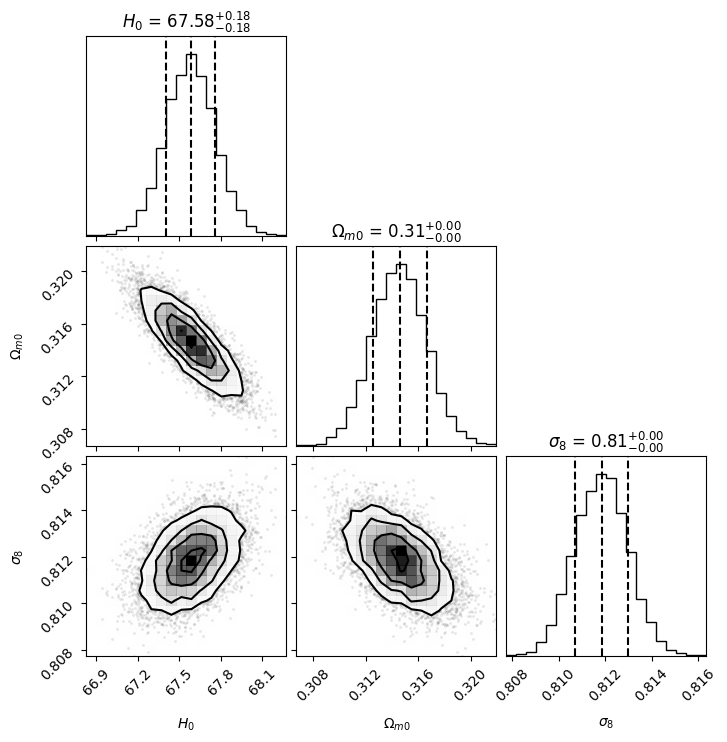}
	}
	\caption{The $68\%$, $95\%$, $99\%$ confidence regions of the joint and marginal posterior probability distributions of $H_0$, $\Omega_{m0}$ and $\sigma_{8}$ that is estimated from our joint constraint method (a) with redshift and redshift error, (b) with redshift but no redshift error, as well as (c) the traditional combined method without redshift.}
	\label{fig:cke}
\end{figure*}

Our methodology can also be applied to these circumstances. For some observations where the redshift errors can be measured, our method can be extended to include the observational information of redshift with error. One can modify the Eq. (\ref{eq8}) to the following expression
\begin{equation}
	\begin{aligned}
\mathcal{T}\left( V_{\mathrm{obs}}^{\mathrm{joint}}\right)&=\bigoplus_{j>k\geqslant 1}\left\lbrace \mathcal{F}_{j}\left(z ; \boldsymbol{\theta}^{j}\right),\mathcal{F}_{k}\left(z ; \boldsymbol{\theta}^{k}\right)\right\rbrace_{\tilde{f}}\\
&+\frac{1}{2} \bigoplus_{j\geqslant 1}\left\lbrace \mathcal{F}_{j}\left(z ; \boldsymbol{\theta}^{j}\right),\mathcal{F}_{j}\left(z ; \boldsymbol{\theta}^{j}\right)\right\rbrace_{\tilde{f}}\\
&+\frac{1}{2}\bigoplus_{j\geqslant 1}\left\lbrace \mathcal{F}^{-1}_{j}\left(z ; \boldsymbol{\theta}^{j}\right),\mathcal{F}^{-1}_{j}\left(z ; \boldsymbol{\theta}^{j}\right)\right\rbrace_{\tilde{f}},
	\end{aligned}
	\label{eq10}
\end{equation}
and the error calculation can be modified correspondingly. Additionally, if there is no redshift observational error and one wishes to use the information of redshift, one can use the following expression
\begin{equation}
	\begin{aligned}
\mathcal{T}\left( V_{\mathrm{obs}}^{\mathrm{joint}}\right)&=\bigoplus_{j>k\geqslant 1}\left\lbrace \mathcal{F}_{j}\left(z ; \boldsymbol{\theta}^{j}\right),\mathcal{F}_{k}\left(z ; \boldsymbol{\theta}^{k}\right)\right\rbrace_{\tilde{f}}\\&+\frac{1}{2} \bigoplus_{j\geqslant 1}\left\lbrace \mathcal{F}_{j}\left(z ; \boldsymbol{\theta}^{j}\right),\mathcal{F}_{j}\left(z ; \boldsymbol{\theta}^{j}\right)\right\rbrace_{\tilde{f}},
	\end{aligned}
	\label{eq11}
\end{equation}
and the error calculation can be modified correspondingly. In addition, our joint constraint method can also be transformed into the form of a traditional combined method to constrain cosmological parameters. One can modify the Eq. (\ref{eq8}) to the following expression
\begin{equation}
	\mathcal{T}\left( V_{\mathrm{obs}}^{\mathrm{combined}}\right)=\frac{1}{2}\bigoplus_{j\geqslant 1}\left\lbrace \mathcal{F}_{j}\left(z ; \boldsymbol{\theta}^{j}\right),\mathcal{F}_{j}\left(z ; \boldsymbol{\theta}^{j}\right)\right\rbrace_{\tilde{f}},
	\label{eq12}
\end{equation}
and the error calculation can also be modified correspondingly. The verification results of parameter constraint of expressions \ref{eq10}, \ref{eq11} and \ref{eq12} are shown in Fig. \ref{fig:cke} respectively. We utilize the mock data generated in Section. \ref{Tsard}, but the difference is that we need to discuss the error of redshift, so we need to generate the error of redshift corresponding to each data point through simulation. We use error transfer to calculate the error of redshift $\sigma_{\tilde{z_j}}$
\begin{equation}
	\sigma_{\tilde{z_j}}=\sum_{i}^{D}\left[ \frac{\partial\mathcal{F}_{j}^{-1}\left(\tilde{f}_j ; \boldsymbol{\theta}^{j}\right)}{\partial \theta^{j}_{i}}\sigma_{\theta^{j}_{i}}\right]^2.
\end{equation}
With mock data, we find that for joint constraint method that use redshift information and have redshift error $\sigma_{\tilde{z_j}}$: $H_0=67.31\pm0.05\mathrm{~km} \mathrm{~s}^{-1} \mathrm{~Mpc}^{-1}$, $\Omega_{m0}=0.3152\pm0.0006$, $\sigma_{8}=0.8110\pm0.0006$ which correspond to expression \ref{eq10}. We find that for the joint constraint method, which uses redshift information but has no redshift error $\sigma_{\tilde{z_j}}$: $H_0=68.9\pm0.1\mathrm{~km} \mathrm{~s}^{-1}\mathrm{~Mpc}^{-1}$, $\Omega_{m0}=0.302\pm0.001$, $\sigma_{8}=0.8137\pm0.0003$ which correspond to expression \ref{eq11}. We find that for combined method that do not use redshift: $H_0=67.6\pm0.2\mathrm{~km} \mathrm{~s}^{-1} \mathrm{~Mpc}^{-1}$, $\Omega_{m0}=0.315\pm0.002$, $\sigma_{8}=0.812\pm0.001$ which correspond to expression \ref{eq12}.

There are several areas in which our method could be improved. Firstly, we encounter some theoretical expressions, such as $f\sigma_{8}(z)$, which cannot be explicitly obtained as an inverse function of redshift. As a compromise solution, we use the segmentally functional expansions shown in Eq. (\ref{eq3}), to find the inverse functions, which increases the complexity of our model. Therefore, it is necessary to explore more efficient methods that can handle such expressions with greater ease. Secondly, we use reconstruction to align the redshift of each observable quantity and generate the observational data, so how well the reconstruction is done directly impacts our final results of constraining cosmological parameters. As mentioned previously, using the reconstructed method to generate data is now a necessary because no observations can capture all the quantities containing cosmological information. And we should find a method that can effectively estimate the variability of the quantity of information in the process of reconstruction, which means that the result of reconstruction has the same  equivalent amount of information to the observational data. Thirdly, when we use different ``move" that is an algorithm for updating the coordinates of walkers in an ensemble sampler based on the current set of coordinates in a manner that satisfies detailed balance. In most instances, the update for each walker is based on the coordinates of the complementary ensemble of walkers. Therefore, we discuss the effect of different ``move"s on the errors of the parameter constraints. We use three different ``move"s: (1) StretchMove as the default for emcee \citep{2010CAMCS...5...65G}, (2) DEMove as runnable under emcee \citep{2014ApJS..210...11N}, and (3) DESnookerMove as runnable under emcee \citep{Cajo:2008sc}. This can be used to get a more efficient sampler for models where the stretch move is not well suited, such as high-dimensional or multi-modal probability surfaces. We use the $H(z)$ side of the mock data generated in Section. \ref{Mocop} for the test, and we listed the results of $H_{0}$ in Table. \ref{table:IV}. As can be seen in the table, the error results vary greatly depending on which of the various "move" modes is selected. On balance, we choose the StretchMove mode of ``move". Overall, addressing these issues can improve the reliability and accuracy of our model and contribute to advancements in cosmological research. Fourthly, composite likelihood has also been used to construct joint distributions in settings where there are no obvious high-dimensional distributions. It is not clear whether or not composite likelihood methods give meaningful results if there is no joint distribution compatible with the component densities used to construct the composite likelihood \citep{23c7e5f7-f5a5-3aad-91bb-4641d35779f8}. And, a major hot potato with maximum likelihood estimation is the difficulty in checking the assumption of multivariate normality \citep{23c7e5f7-f5a5-3aad-91bb-4641d35779f8}. Therefore, one can try to construct a certain distribution to overcome the problem.
\begin{table*}[!ht]
	\centering
	\tabcolsep=0.1cm
	\caption{The error of $H_0$ constraints under different MCMC ``move".}
		\begin{tabular}{cccccccc}
			\hline\hline StretchMove weight*& DEMove weight &DESnookerMove weight&Run $\text{I}^{\star}$&Run II&Run III&Run IV&Run V\\
		\hline	$100\%$&$0\%$&$0\%$&$(0.25,0.25)$&$(0.25,0.25)$&$(0.25,0.24)$&$(0.25,0.25)$&$(0.24,0.25)$\\
			$0\%$&$100\%$&$0\%$&$(0.25,0.25)$&$(0.25,0.25)$&$(0.24,0.25)$&$(0.24,0.25)$&$(0.24,0.24)$\\
			$0\%$&$0\%$&$100\%$&$(0.15,0.15)$&$(0.15,0.15)$&$(0.15,0.15)$&$(0.15,0.15)$&$(0.15,0.15)$\\
			$0\%$&$10\%$&$90\%$&$(0.16,0.17)$&$(0.16,0.17)$&$(0.16,0.17)$&$(0.16,0.17)$&$(0.16,0.17)$\\
			$0\%$&$20\%$&$80\%$&$(0.18,0.18)$&$(0.18,0.19)$&$(0.17,0.18)$&$(0.18,0.18)$&$(0.17,0.17)$\\
			$0\%$&$30\%$&$70\%$&$(0.19,0.19)$&$(0.20,0.19)$&$(0.20,0.19)$&$(0.19,0.20)$&$(0.20,0.19)$\\
			$0\%$&$40\%$&$60\%$&$(0.21,0.19)$&$(0.21,0.21)$&$(0.20,0.20)$&$(0.20,0.21)$&$(0.20,0.20)$\\
			$0\%$&$50\%$&$50\%$&$(0.21,0.21)$&$(0.20,0.21)$&$(0.20,0.20)$&$(0.21,0.20)$&$(0.20,0.21)$\\
			$0\%$&$60\%$&$40\%$&$(0.21,0.23)$&$(0.21,0.22)$&$(0.20,0.21)$&$(0.21,0.22)$&$(0.21,0.21)$\\
			$0\%$&$70\%$&$30\%$&$(0.22,0.21)$&$(0.22,0.23)$&$(0.23,0.22)$&$(0.21,0.23)$&$(0.21,0.24)$\\
			$0\%$&$80\%$&$20\%$&$(0.22,0.22)$&$(0.23,0.23)$&$(0.23,0.23)$&$(0.22,0.23)$&$(0.22,0.24)$\\
			$0\%$&$90\%$&$10\%$&$(0.24,0.23)$&$(0.22,0.23)$&$(0.22,0.24)$&$(0.23,0.23)$&$(0.24,0.23)$\\
			\hline
		\end{tabular}
	\begin{flushleft}
\tablecomments{
* Normalized weight that is used to update the coordinates of each walker that ``moves" with the weighted probability.\\
${}^{\star}$ The upper and lower error pairs for $H_{0}$ constraints from five runs of the MCMC chain completely after removing the unstable chain before burn-in.}
\end{flushleft}
	\label{table:IV}
\end{table*}

\section{Conclusion}
\label{conclusion}
In this paper, we propose a novel joint cosmological constraint method that eliminates the redshift by exploiting the interdependence of multiple observables from the same object or at the same redshift $z$. We begin by validating the efficacy and capacity of the cosmological model to constrain parameters with a set of carefully simulated data. As it is not possible to observe all the observed space $V_{\mathrm{obs}}^{\mathrm{joint}}$ for a certain object at the same time, we regard the reconstructed observables and data at different redshifts within overlapping redshift range as the ``real" observational data to obtain the results of our joint constraint method. We summarize our findings below. (1) Redshift is no longer a mapping corresponding to theoretical and observational values, hence the uncertainty of data caused by redshift no longer exists. (2) Comparing with the traditional combined parameter inference method, our method makes maximum use of the information in the observational data, thus reducing the error of the constraint results and having the advantage of breaking the parametric degeneracies to a certain extent. (3) The precision of our implicit redshift system is already very high, and the model error is significantly less than the data error through calculation. Consequently, it can be considered that the error of parameter inference originates from the errors of observational data and reconstruction. At the same time, because of the high precision of our system, the model complexity of our method is significantly greater than that of the traditional combined model with parameter constraints. (4) Because we make maximum use of the observational data by Eq. (\ref{eq5}), we can use fewer observational data to obtain the results of cosmological constraints than the traditional method, albeit with larger errors than in the case of using the whole dataset. (5) For some observations where the redshift errors can be measured, our method can be extended to include the observational information of redshift with error. Moreover, if there is no observational error of redshift, but one wants to use the information of redshift, our method still works. In addition, our joint constraint method can also be degraded to the form of a traditional combined method to constrain cosmological parameters. 

In the future, we plan to find a more concise and efficient methods to solve the inverse function to reduce the complexity of our model, with the aim of reducing the complexity of our model. One such approach is deep learning, which has shown promise in solving complex mathematical problems. Furthermore, we intend to consider better methods to directly write an analytical expression of multiple observables, such as using the generalized function $\Lambda_{\mu}\left(\mathcal{F}_{1},\mathcal{F}_{2},\cdots,\mathcal{F}_{M} \right)$ from functional analysis, which have two reasons to do this: firstly, one can save the computational complexity of repeated solutions from inverse function class $\tilde{z_j}$. Secondly, a major reason for concern with maximum likelihood estimation is the difficulty in checking the assumption of multivariate normality. And a certain distribution function can solve the problem. Moreover, we also intend to come up with a way to measure as many observables of the same object as possible containing cosmological information such as BAO that both the Hubble expansion parameter and the angular diameter distance at the same redshift can be obtained in one observation. In this manner, errors in data reconstruction can be eliminated and parametric degeneracies can be attenuated, which we discuss in Section. \ref{Tsard}.

\begin{acknowledgments}
We are grateful for the referee's insightful and useful comments, which helped us improve our manuscript. T.-J.Z. (张同杰) dedicates this paper to the memory of his mother, Yu-Zhen Han (韩玉珍), who passed away 3 yr ago (2020 August 26). We are grateful to Jing Niu, Jie-feng Chen, Xiao-Hang Luan and Jian-Kang Li for useful discussions. This work was supported by the National Science Foundation of China (Grants No. 61802428, 11929301).
\end{acknowledgments}

\software{Python, ReFANN \citep{2020ApJS..246...13W},
	emcee \citep{2013PASP..125..306F},
	NumPy \citep{2020Natur.585..357H},
	pandas \citep{2023zndo...7794821T,mckinney-proc-scipy-2010},
	Matplotlib \citep{2007CSE.....9...90H}
}

\appendix
\label{appendix}
\section{The Method of Alleviating Model's Tensions}
\begin{figure}[htbp]
	\centering
	\subfigure[]{
		\includegraphics[width=0.33\linewidth]{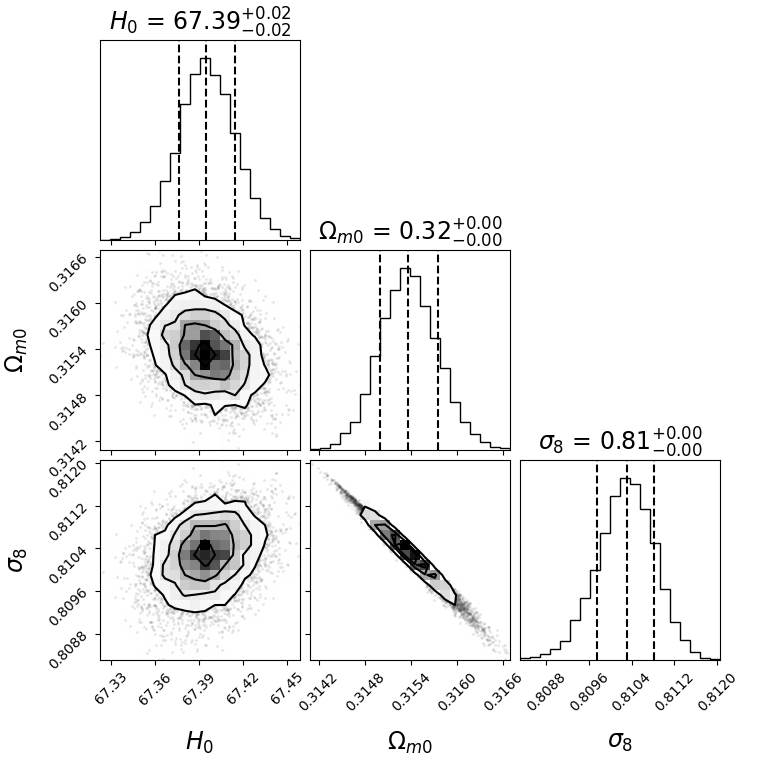}
	}
	\subfigure[]{
		\includegraphics[width=0.33\linewidth]{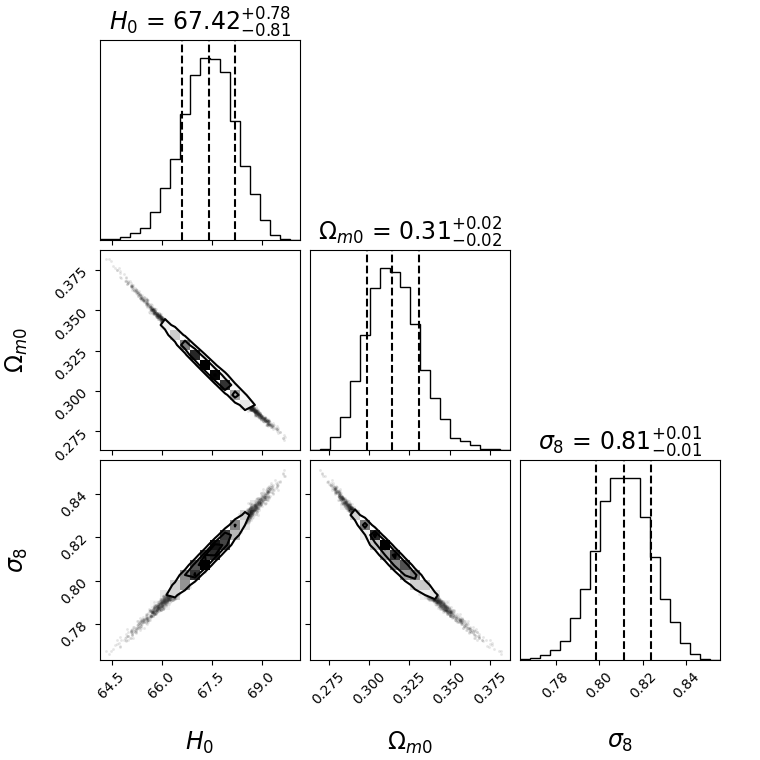}
	}
	\subfigure[]{
		\includegraphics[width=0.33\linewidth]{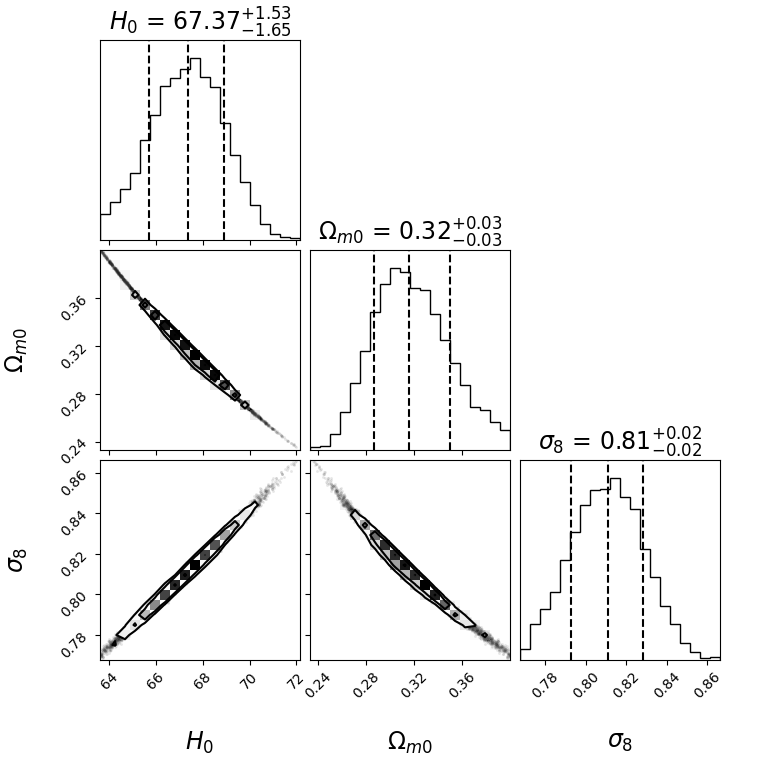}
	}
	\subfigure[]{
	\includegraphics[width=0.33\linewidth]{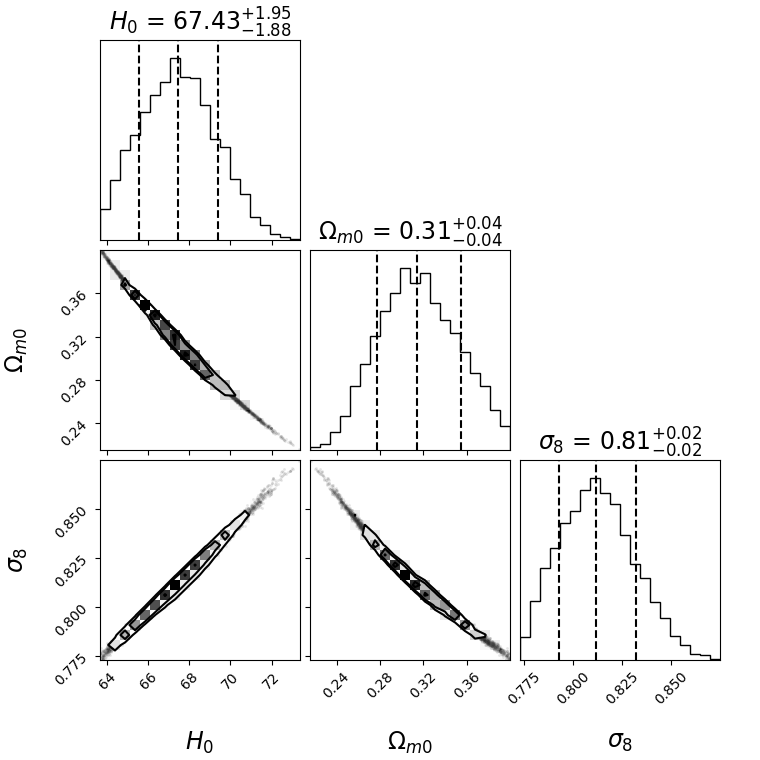}
	}
	\caption{The $68\%$, $95\%$, $99\%$ confidence regions of the joint and marginal posterior probability distributions of $H_0$, $\Omega_{m0}$ and $\sigma_{8}$ that is estimated from our joint constraint method with the interval of the redshift bin (a) $z\in[0,0.8]$, (b) $z\in[0.8,1.5]$, (c) $z\in[1.5,2.3]$, as well as (d) $z\in[2.3,4]$.}
	\label{appendix-bin-mock}
\end{figure}

It is widely recognized that numerous cosmological models encounter tensions when formulated at both high and low redshifts. In order to address this issue, numerous researchers are actively exploring cosmographic methodologies that can effectively reconcile early and late epochs \citep{2020MNRAS.494.2576C,2022MNRAS.509.5399C,2021CQGra..38r4001K,OCOLGAIN2023101216}. The $\Lambda$CDM model is largely only extensively examined in the dark energy dominated regime ($z \lesssim1$) and at significantly high redshifts ($z\sim1100$). However, there remains a considerable range of redshifts that have yet to be investigated in order to validate or challenge the model. According to the redshift bin method mentioned in \citep{OCOLGAIN2023101216} and combining it with the redshift interval of our mock data generated when verifying formula 8, we divide the data into four bins, with sub-bins defined by the intervals [0,0.8], [0.8,1.5], [1.5,2.3], and [2.3,4]. In order to constrain cosmological parameters, we make the assumption that the priors for $H_0$, $\Omega_{m0}$, and $\sigma_{8}$ are within the ranges of $[0,100]$, $[0,1]$, and $[0,1]$ respectively. Therefore, we find that (a) at sub-bin $z\in[0,0.8]$: $H_0=67.39\pm0.02 \mathrm{~km} \mathrm{~s}^{-1} \mathrm{~Mpc}^{-1}$, $\Omega_{m0}=0.3154\pm0.0004$, $\sigma_{8}=0.8103\pm0.0006$. (b) At sub-bin $z\in[0.8,1.5]$: $H_0=67.4\pm0.8 \mathrm{~km} \mathrm{~s}^{-1} \mathrm{~Mpc}^{-1}$, $\Omega_{m0}=0.31\pm0.02$, $\sigma_{8}=0.81\pm0.01$. (c) At sub-bin $z\in[1.5,2.3]$:    $H_0=67.4^{+1.5}_{-1.7} \mathrm{~km} \mathrm{~s}^{-1} \mathrm{~Mpc}^{-1}$, $\Omega_{m0}=0.32\pm0.03$, $\sigma_{8}=0.81\pm0.02$. (d) At sub-bin $z\in[2.3,4]$: $H_0=67.4\pm1.9\mathrm{~km} \mathrm{~s}^{-1} \mathrm{~Mpc}^{-1}$, $\Omega_{m0}=0.31\pm0.04$, $\sigma_{8}=0.81\pm0.02$. We show the results in Fig. \ref{appendix-bin-mock}. Interestingly, in the case of our mock data, we find that the Hubble constant $H_0$ does not evolve with redshift in the redshift interval as mentioned in \citep{OCOLGAIN2023101216}. This intriguing phenomenon has been observed in two distinct datasets, each analyzed independently.

\begin{figure}[htbp]
	\centering
	\subfigure[]{
		\includegraphics[width=0.31\linewidth]{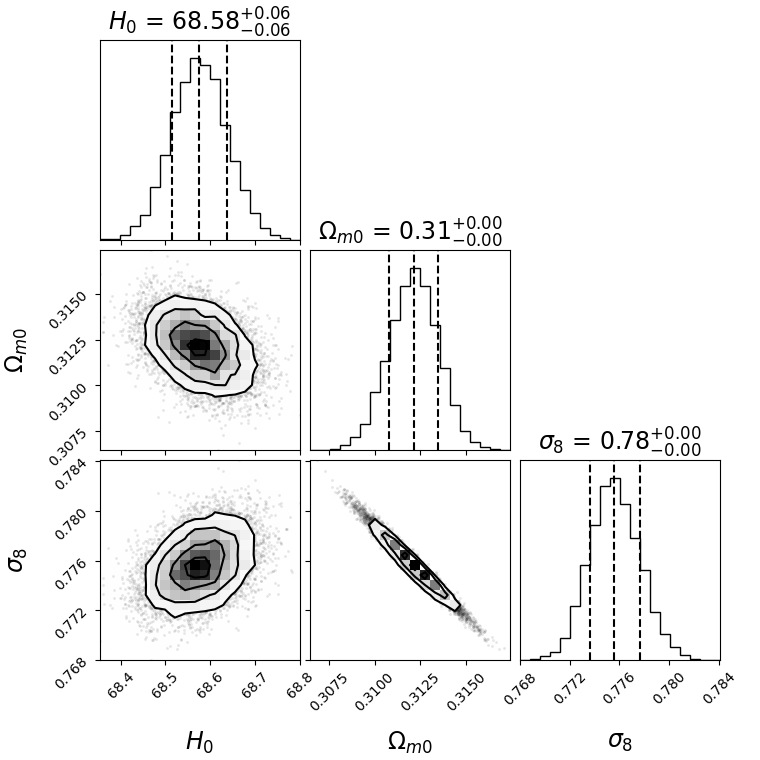}
	}
	\subfigure[]{
		\includegraphics[width=0.31\linewidth]{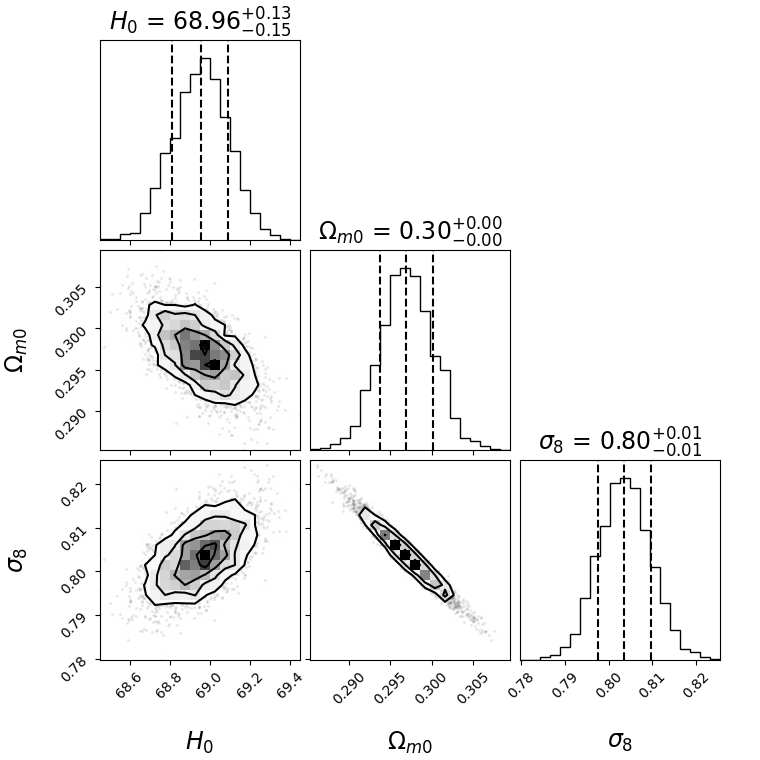}
	}
	\subfigure[]{
		\includegraphics[width=0.31\linewidth]{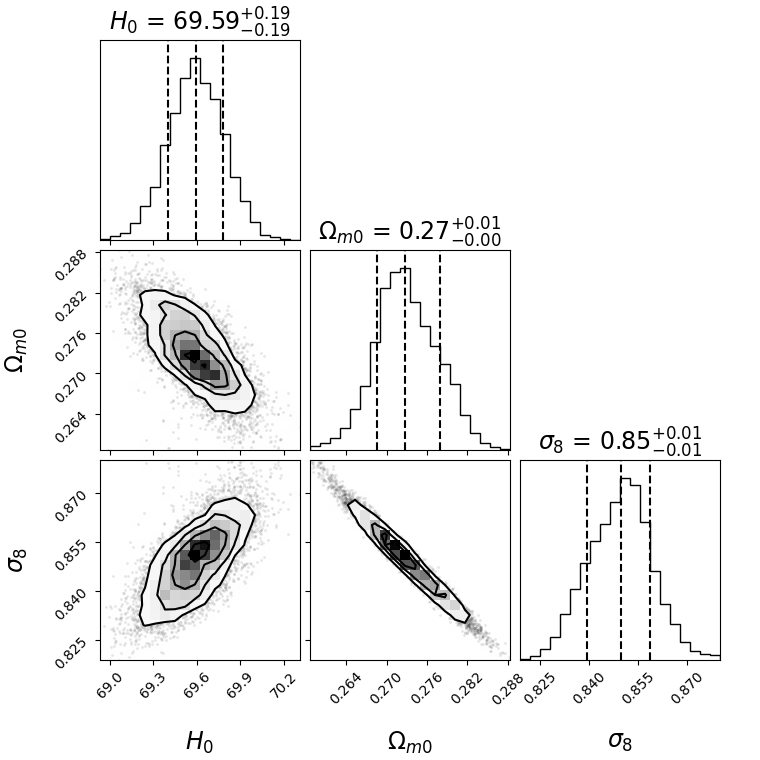}
	}
	\caption{The $68\%$, $95\%$, $99\%$ confidence regions of the joint and marginal posterior probability distributions of $H_0$, $\Omega_{m0}$ and $\sigma_{8}$ that is estimated from our joint constraint method with the interval of the redshift bin (a) $z\in[0.11,0.8]$, (b) $z\in[0.8,1.5]$, as well as (c) $z\in[1.5,1.944]$.}
	\label{appendix-bin-rec}
\end{figure}

However, this interesting phenomenon is not observed when we use the reconstructed data, and the Hubble constant $H_0$ continues to evolve with redshift. Considering the different redshift range of the data, we divide the reconstructed data into three bins, and the interval of the sub-bin is $[0.11,0.8]$, $[0.8,1.5]$ and $[1.5,1.944]$. We assume the priors $H_0\in[0,100]$, $\Omega_{m0}\in[0,1]$ and $\sigma_{8}\in[0,1]$ to constrain cosmological parameters. Hence, we find that (a) at sub-bin $z\in[0.11,0.8]$: $H_0=68.6\pm0.1 \mathrm{~km} \mathrm{~s}^{-1} \mathrm{~Mpc}^{-1}$, $\Omega_{m0}=0.312\pm0.001$, $\sigma_{8}=0.776\pm0.002$. (b) At sub-bin $z\in[0.8,1.5]$: $H_0=69.0^{+0.1}_{-0.2} \mathrm{~km} \mathrm{~s}^{-1} \mathrm{~Mpc}^{-1}$, $\Omega_{m0}=0.297\pm0.003$, $\sigma_{8}=0.804\pm0.006$. (c) At sub-bin $z\in[1.5,2.3]$:    $H_0=69.6\pm 0.2\mathrm{~km} \mathrm{~s}^{-1} \mathrm{~Mpc}^{-1}$, $\Omega_{m0}=0.273^{+0.005}_{-0.004}$, $\sigma_{8}=0.85\pm 0.01$. We show the results in Fig. \ref{appendix-bin-rec}.

We believe that there are factors that contribute to this outcome. (1) As discussed above, when we reconstruct the data we need, we need to find a way to measure the changes in the cosmological information in the observational data during this process. It is conceivable that the cosmological information may have undergone modifications during the reconstruction process, resulting in different Hubble constant constraining results for different redshift bin intervals. For mock data, it is noteworthy to observe that the aforementioned statement does not hold true, which presents an ``interesting phenomenon". This is because every time we input cosmological parameters from Planck, the resulting data is tested for goodness of fit to ensure that it conforms to the corresponding Gaussian distribution with the rejection region $W=\left\lbrace \chi^2>\chi^2_{0.95} \right\rbrace$. (2) The available observational data do not inherently guarantee the consistency of cosmological parameter constraints over each redshift interval. The $\Lambda$CDM model is largely only well tested at low redshifts $z \lesssim1$ and at very high redshifts $z\sim1100$. A recent article using the JWST's high redshift observations $(z\sim15)$ indicates that there is a great tension between the JWST results and the $\Lambda$CDM model \citep{2023MNRAS.524.3385G}. In conclusion, our method can make an attempt in this matter, especially after determining the variation of cosmological information in the reconstruction methodology.

\section{The Method of Alleviating $H_0$ Tensions}
\begin{figure}[htbp]
	\centering
	\subfigure[]{
		\includegraphics[width=0.33\linewidth]{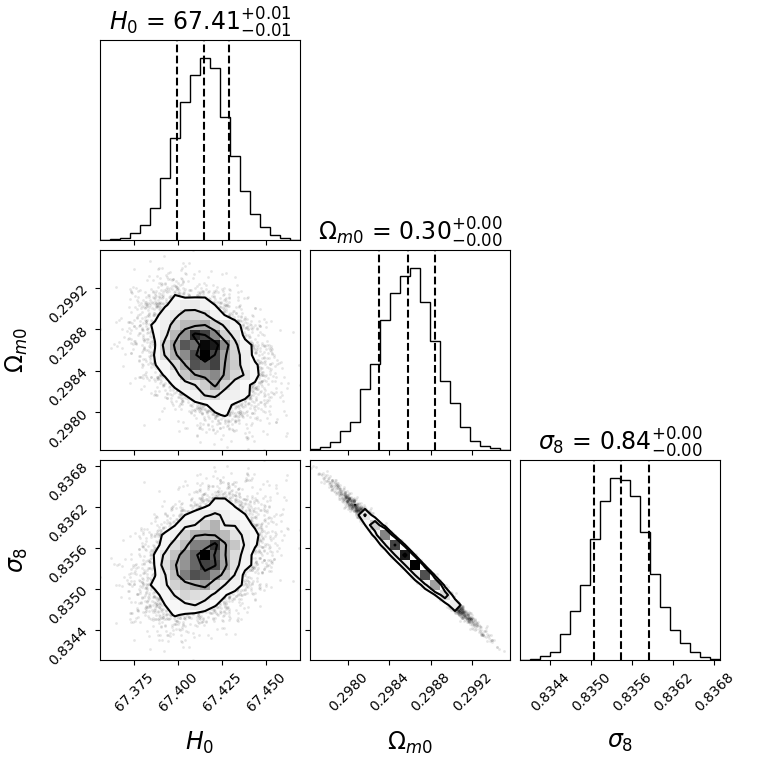}
	}
	\subfigure[]{
		\includegraphics[width=0.33\linewidth]{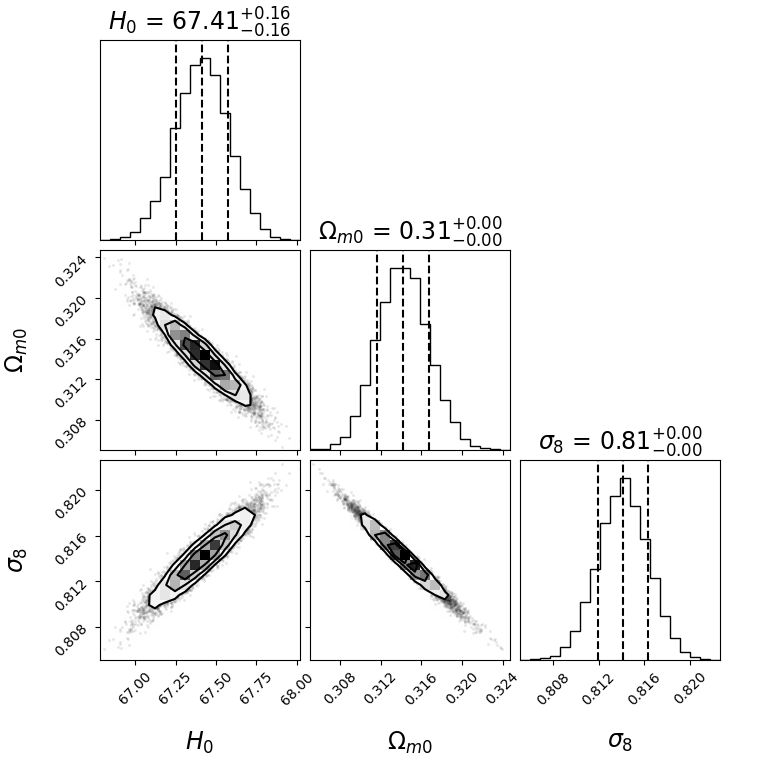}
	}
	\subfigure[]{
		\includegraphics[width=0.33\linewidth]{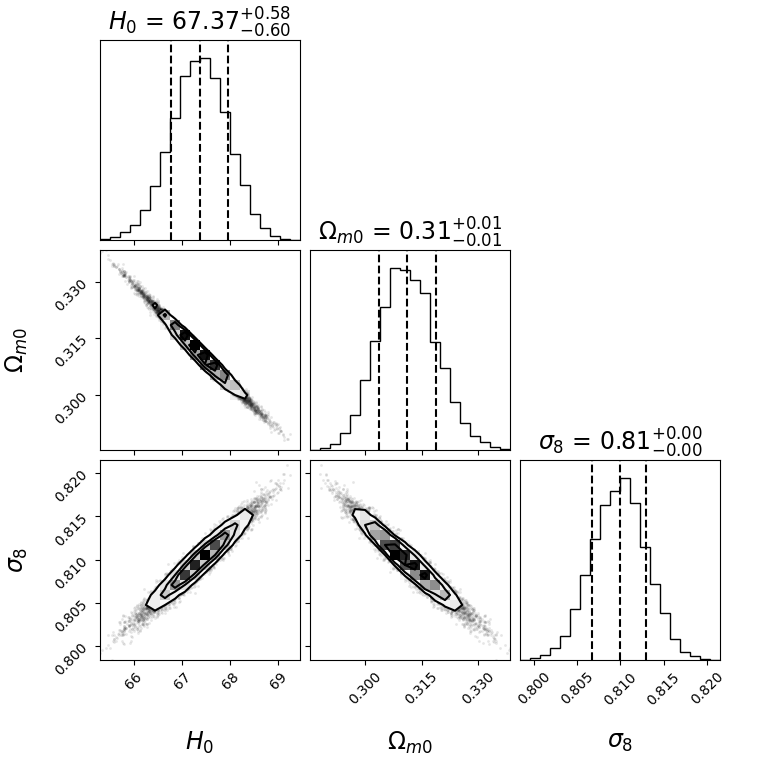}
	}
	\subfigure[]{
		\includegraphics[width=0.33\linewidth]{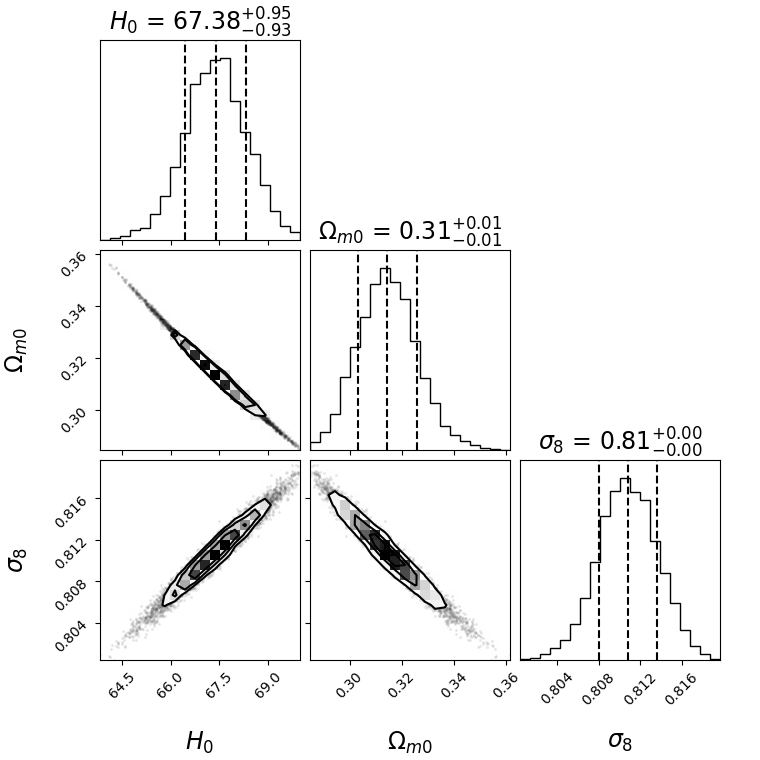}
	}
	\caption{The $68\%$, $95\%$, $99\%$ confidence regions of the joint and marginal posterior probability distributions of $H_0$, $\Omega_{m0}$ and $\sigma_{8}$ that is estimated from our joint constraint method with the interval of the redshift bin (a) $z\in[0,0.8]$, (b) $z\in[0.8,1.5]$, (c) $z\in[1.5,2.3]$, as well as (d) $z\in[2.3,4]$.}
	\label{appendix-bin-dl-mock}
\end{figure}

It appears that the tension in the field of cosmology is closely associated with the phenomenon of redshift. The current methodology has the potential to effectively address this particular issue. For instance, it appears that the tension associated with the $H_0$ might potentially be mitigated or alleviated through modifications to the cosmological observables $\mathcal{F}_{i,\mathrm{obs}}$, as discussed in the works of \citep{2023PDU....4001201C} and \citep{2021MNRAS.502.2065D}. We change the cosmological observables used from angular diameter $D_A(z)$ to luminosity distance $D_L(z)$ for the test. According to the redshift bin method mentioned in \citep{OCOLGAIN2023101216} and combining it with the redshift interval of our mock data generated when verifying formula 8 but with new data about luminosity distance $D_L(z)$, we divide it into four bins, and the interval of the sub-bin is $[0,0.8]$, $[0.8,1.5]$, $[1.5,2.3]$ and $[2.3,4]$. We assume the priors $H_0\in[0,100]$, $\Omega_{m0}\in[0,1]$ and $\sigma_{8}\in[0,1]$ to constrain cosmological parameters. Therefore, we find that (a) at sub-bin $z\in[0,0.8]$: $H_0=67.41\pm0.01 \mathrm{~km} \mathrm{~s}^{-1} \mathrm{~Mpc}^{-1}$, $\Omega_{m0}=0.2986\pm0.0003$, $\sigma_{8}=0.8354\pm0.0004$. (b) At sub-bin $z\in[0.8,1.5]$: $H_0=67.4\pm0.2 \mathrm{~km} \mathrm{~s}^{-1} \mathrm{~Mpc}^{-1}$, $\Omega_{m0}=0.31\pm0.03$, $\sigma_{8}=0.81\pm0.01$. (c) At sub-bin $z\in[1.5,2.3]$: $H_0=67.4\pm0.6 \mathrm{~km} \mathrm{~s}^{-1} \mathrm{~Mpc}^{-1}$, $\Omega_{m0}=0.311^{+0.008}_{-0.007}$, $\sigma_{8}=0.810\pm0.003$. (d) At sub-bin $z\in[2.3,4]$: $H_0=67.4\pm0.9\mathrm{~km} \mathrm{~s}^{-1} \mathrm{~Mpc}^{-1}$, $\Omega_{m0}=0.31\pm0.01$, $\sigma_{8}=0.811\pm0.003$. We show the results in Fig. \ref{appendix-bin-dl-mock}. Interestingly, in the case of our mock data, we also find that the Hubble constant $H_0$ does not evolve with redshift in the redshift interval as mentioned in \citep{OCOLGAIN2023101216}.

In conjunction with the discussion in the previous appendix, it is not surprising that this result occurs. Our mock data comes from Planck's corresponding $\Lambda$CDM model, which is compliant with the $\Lambda$CDM model without any tension-inducing. The generated mock data exhibits minor oscillations rather than substantial inconsistencies between high and low redshift outcomes. In conclusion, our approach demonstrates a modest but noteworthy endeavor in this particular domain, particularly following a careful assessment of the tension-inducing data within the $\Lambda$CDM framework.

\bibliography{ancjcimoswri}
\bibliographystyle{aasjournal}

\end{CJK*}
\end{document}